\documentclass[aps, twocolumn, secnumarabic, amssymb, nobibnotes, prl, superscriptaddress, 10pt]{revtex4-2}

\newcommand{\murm}{
  \ifmmode
    \mathchoice
        {\hbox{\normalsize\textmu}}
        {\hbox{\normalsize\textmu}}
        {\hbox{\scriptsize\textmu}}
        {\hbox{\tiny\textmu}}
  \else
   ~\textmu
  \fi
}

\usepackage{bbold}

\usepackage{lipsum}
\usepackage{amsfonts,amssymb,amsmath}
\usepackage{graphicx}
\usepackage{bm}
\usepackage{calc}
\usepackage[x11names]{xcolor}
\usepackage{siunitx}
\usepackage{tikz}
\usepackage{wasysym}
\usepackage{dsfont}
\usepackage{hyperref}
\usepackage{diagbox}

\newcommand{\fmarki}{*}
\newcommand{\fmarkii}{\ensuremath{\dagger}}
\newcommand{\fmarkiii}{\ensuremath{\ddagger}}

\def\@fnsymbol#1{{\ifcase#1\or \fmarki\or \fmarkii\or \fmarkiii\or \fmarkiv\or \fmarkv\or \fmarkvi\or \fmarkvii\or \fmarkviii\or \fmarkix \else\@ctrerr\fi}}
\makeatother

\renewcommand{\fmarki}{b$_1$}
\renewcommand{\fmarkii}{b$_2$}
\renewcommand{\fmarkiii}{c$_3$}

\newcommand{\nvn}{N-$V$}
\newcommand{\nv}{N-$V$\,}
\newcommand{\nuc}{${}^{13}$C }
\newcommand{\nucn}{${}^{13}$C}
\newcommand{\ctw}{${}^{12}$C }

\makeatletter
\renewcommand\subsubsection[1]{
  \refstepcounter{subsubsection}
  \par
  \vskip 1.0\baselineskip 
  \begin{center}
    {\normalfont\small\bfseries \thesubsubsection\ #1}
  \end{center}
  \vskip 0.5\baselineskip 
}
\makeatother

\begin{document}

\title{Probing Many-Body Phenomena with Atomically Thin Nuclear Spin Layers in Diamond}

\author{Philipp J. Vetter}
\email{philipp.vetter(at)uni-ulm.de}
\affiliation{Institute for Quantum Optics, Ulm University, Albert-Einstein-Allee 11, 89081 Ulm, Germany}
\affiliation{Center for Integrated Quantum Science and Technology (IQST), 89081 Ulm, Germany}
\author{Christoph Findler}
\affiliation{Institute for Quantum Optics, Ulm University, Albert-Einstein-Allee 11, 89081 Ulm, Germany}
\affiliation{Center for Integrated Quantum Science and Technology (IQST), 89081 Ulm, Germany}
\affiliation{Diatope GmbH, Buchenweg 23, 88444 Ummendorf, Germany}
\author{Antonio Verdú}
\affiliation{Departamento de Física, Universidad de Murcia, 30071 Murcia, Spain}
\author{Matthias Kost}
\affiliation{Institute of Theoretical Physics, Ulm University, Albert-Einstein-Allee 11, 89081 Ulm, Germany}
\affiliation{Center for Integrated Quantum Science and Technology (IQST), 89081 Ulm, Germany}
\author{Rémi Blinder}
\affiliation{Institute for Quantum Optics, Ulm University, Albert-Einstein-Allee 11, 89081 Ulm, Germany}
\affiliation{Center for Integrated Quantum Science and Technology (IQST), 89081 Ulm, Germany}
\author{Jens Fuhrmann}
\affiliation{Institute for Quantum Optics, Ulm University, Albert-Einstein-Allee 11, 89081 Ulm, Germany}
\affiliation{Center for Integrated Quantum Science and Technology (IQST), 89081 Ulm, Germany}
\author{Christian Osterkamp}
\affiliation{Diatope GmbH, Buchenweg 23, 88444 Ummendorf, Germany}
\author{Johannes Lang}
\affiliation{Diatope GmbH, Buchenweg 23, 88444 Ummendorf, Germany}
\author{Martin B. Plenio}
\affiliation{Institute of Theoretical Physics, Ulm University, Albert-Einstein-Allee 11, 89081 Ulm, Germany}
\affiliation{Center for Integrated Quantum Science and Technology (IQST), 89081 Ulm, Germany}
\author{Javier Prior}
\affiliation{Departamento de Física, Universidad de Murcia, 30071 Murcia, Spain}
\author{Fedor Jelezko}
\email{fedor.jelezko(at)uni-ulm.de}
\affiliation{Institute for Quantum Optics, Ulm University, Albert-Einstein-Allee 11, 89081 Ulm, Germany}
\affiliation{Center for Integrated Quantum Science and Technology (IQST), 89081 Ulm, Germany}
%



\begin{abstract}
Quantum simulation aims to recreate complex many-body phenomena in controlled environments, offering insights into dynamics that are otherwise difficult to model.
Existing platforms, however, are often complex and costly to scale, typically requiring ultra pure vacuum or low temperatures.
Here, we introduce a platform based on a thin, strongly interacting \nuc nuclear spin layer in diamond that allows controlled exploration of many-body dynamics at room temperature.
Nearby nitrogen-vacancy centers enable polarization, readout, and, combined with radio-frequency fields, coherent control of the nuclear spins.
We demonstrate strong, tunable interactions among the nuclear spins and use the system to probe discrete time-crystalline order across varying interaction ranges. 
By combining ease of use with operation at ambient temperatures, our work opens new opportunities for investigating strongly correlated many-body effects.
\end{abstract}

\maketitle



%
Despite a century of progress, large-scale quantum phenomena remain difficult to study~\cite{schrodinger2025mechanics}.
Quantum simulators, systems that recreate these complex dynamics in controlled environments, allow one to explore such phenomena directly.
Recent achievements include simulating cosmological effects~\cite{viermann2022quantum, maceda2025digital}, investigating chemical dynamics~\cite{richerme2023quantum, navickas2025experimental}, exploring quantum chromodynamics~\cite{banuls2020simulating}, and creating new phases of matter~\cite{semeghini2021probing, randall2021many, mi2022time, choi2017observation}. 
Most platforms, however, require demanding conditions, such as low temperatures or ultra pure vacuum, raising the following question: Can more accessible systems probe similar dynamics? \\ 
Nitrogen-vacancy (\nvn) centers in diamond offer optical readout and long coherence times at room temperature~\cite{doherty2013nitrogen}, and have been used to study many-body phenomena~\cite{choi2017observation, gao2025signal, wu2025spin}.  
Hyperfine coupling to surrounding nuclear spins allows these nuclei to be used as a resource for quantum simulation, provided their interactions are sufficiently strong, i.e., the nuclear spins are densely packed~\cite{cai2013large}.
A dense nuclear spin layer offers the added advantage of forming a self-assembled, spatially ordered spin register that is partially programmable~\cite{randall2021many}, unlike dense \nv center ensembles where disorder and spectral crowding limit controllability.
Past efforts to create such dense nuclear spin layers involved surface termination~\cite{cai2013large}, depositing two-dimensional materials on the diamond~\cite{lovchinsky2017magnetic}, and delta doping~\cite{unden2018coherent}. 
None of these systems have yet demonstrated their applicability to probe many-body interactions, due to either rapid depolarization of the nuclear spins or the absence of strong dipolar coupling among them. \\
In this work, we present a different approach, based on isotopic engineering of diamond through indirect overgrowth~\cite{findler2020indirect}.
We create a thin \nuc nuclear spin layer ($\le 1$\,nm) next to single \nv centers, embedded in a spin-free \ctw environment, where the \nv centers are positioned via low-energy ${}^{15}\text{N}^+$ implantation during intermediate overgrowth steps. 
Our method preserves the coherence of both \nv centers and nuclear spins, a major improvement over approaches requiring near-surface \nv centers. \\
We estimate the distance between \nv centers and \nuc layer using a numerical model based on Pauli string truncation, and initialize and read out the nuclear spin layer via the \nv centers. 
Probing the \nuc layer’s coherence properties reveals strong, tunable interactions among the nuclear spins, which we harness to demonstrate the system’s potential through the generation and study of discrete time-crystalline order~\cite{yao2017discrete}.
By enabling strong, controllable spin interactions at room temperature, this platform provides a scalable route for exploring strongly correlated quantum phases.

\section{Sample Fabrication and Characterization}
\begin{figure}
    \centering
    \includegraphics[width=0.5\textwidth]{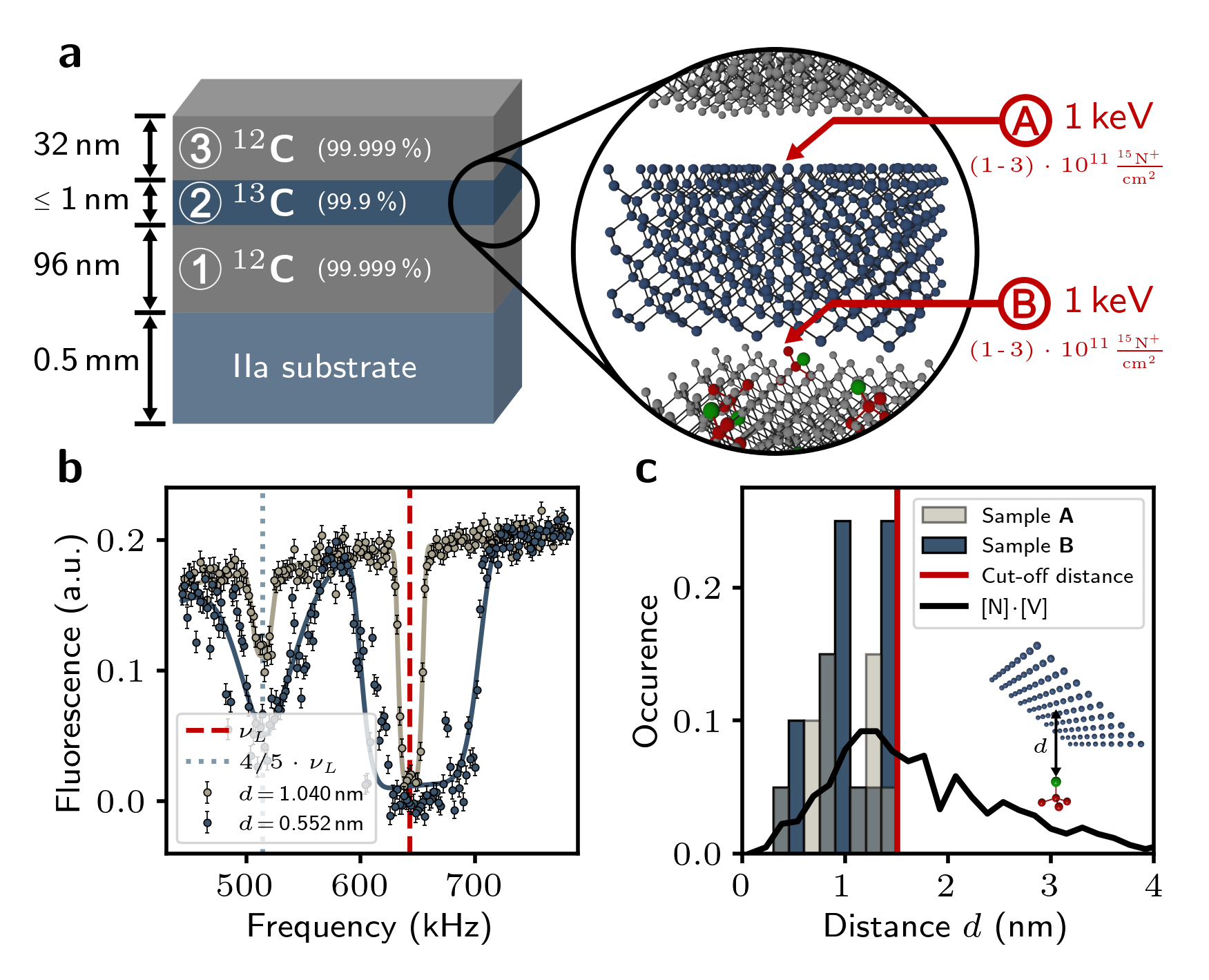}
    \caption{Sample fabrication and characterization.
    \textbf{a} Structure of our diamond samples with numbered overgrowth steps; layer colors indicate the carbon isotope and its corresponding isotopic purity.
    The close-up shows an accurate representation of the atomic structure of the \nuc layer and the red arrows highlight the implantation steps for sample A and B. 
    \textbf{b} AXY spectra of \nv centers with different distances to the nuclear spin layer.
    The spectra show two dips around the Larmor frequency (red dashed line) and a higher harmonic (dark blue dashed line), with distance-dependent linewidths. 
    \textbf{c} Distance distribution for both diamond samples up to the estimation limit, marked by the red line.
    The black line represents the product of the implantation-induced vacancy density [V] and nitrogen density [N].
    The inset visualizes the relative angle of the \nuc layer to the \nv center's symmetry axis.}
    \label{fig:fig1}
\end{figure}
To realize our quantum simulator in a controlled and reproducible manner, we fabricate and characterize two diamond samples. 
Sample A is implanted \textbf{a}fter the \nuc layer growth, while sample B is implanted \textbf{b}efore (see \autoref{fig:fig1}\,a).
We begin by indirect overgrowth~\cite{findler2020indirect} of a commercial diamond substrate with 96\,nm \ctw ($I=0$) in our home-built chemical vapor deposition (CVD) reactor~\cite{silva2010microwave, osterkamp2021engineering}.
Sample B is then implanted with 1\,keV of ${}^{15}\text{N}^+$ to create shallow, single \nv centers~\cite{lang2023atomic}. 
Both samples are subsequently overgrown with a thin \nuc ($I=1/2$) layer of  $\le 1$\,nm thickness after which sample A is implanted. 
Finally, a 32\,nm \ctw cap layer is grown to protect the spins from surface noise. 
The samples are annealed at 1000\,${}^{\circ}$C for 3\,h in a ultra-high vacuum oven to form \nv centers from the implanted nitrogen.
A detailed, step-by-step description of the growth process is provided in the Appendix. 
Due to the [100] growth, the \nuc layer is tilted by 54.7${}^{\circ}$ with respect to the \nv center's symmetry axis. \\
To analyze the growth process, we probe the \nv center's local \nuc environment with the adaptive XY (AXY) dynamical decoupling sequence~\cite{casanova2015robust}.
By replacing the $\pi$ pulses of standard dynamical decoupling sequences with a composite pulse, the AXY sequence allows tuning of the interaction strength between the NV center and surrounding nuclear spins, enabling resolution of distant spins in the presence of close, strongly coupled ones~\cite{casanova2015robust}.
Each nuclear spin creates a resonance that is shifted from the bare Larmor frequency $\nu_L$ by its respective hyperfine coupling~\cite{taminiau2012detection}.
Instead of individual resonances, the large number of \nuc nuclei leads to one broad dip around $\nu_L$ in the AXY spectra in \autoref{fig:fig1}\,b, whose width depends on the distance of the \nv center to the \nuc layer. 
Finite pulse lengths combined with strong $A_{zx}$ and $A_{zy}$ hyperfine interaction lead to an additional harmonic at $4/5\cdot\nu_L$~\cite{loretz2015spurious}. \\
All experiments are performed at room temperature in a confocal setup, with microwave and rf control applied via a simple copper wire, placed on the diamond's surface~\cite{supplement}. 
The magnetic field of $B=600\,\text{G}$ is aligned along the \nv center's symmetry axis. \\ 
Using a truncated subspace method in a Pauli string basis, we simulate the measured spectra with a two-dimensional grid of 100 nuclear spins at varying distances from the \nv center (inset, \autoref{fig:fig1} c).
With an appropriate truncation~\cite{schaetzle2024partitioncce}, we can prioritize Pauli strings of low weight and factor in the different interaction strength magnitudes between the spin species. 
Our model incorporates any two-spin correlations and all three-spin correlations that involve the \nv center~\cite{supplement}. 
Equivalent to the experiment, the spectra are fitted with a generalized normal distribution~\cite{nadarajah2005generalized}, allowing us to correlate the variance with the distance.
Because the signal strength scales with the hyperfine interaction, which decreases with increasing distance, it vanishes beyond $1.5\,\text{nm}$ ("cut-off distance") for the chosen AXY settings. \\ 
For each sample, we measure the AXY spectrum for 20 different \nv centers and plot the estimated distances as a histogram in \autoref{fig:fig1}\,c.
Within the measurable range, both samples show a mean distance of $1\,\text{nm}$~\cite{supplement}. 
For sample A, $50\,\%$ of the \nv centers fall within this range, compared to $90\,\%$ for sample B.
This is counterintuitive, since sample A is implanted after the \nuc overgrowth, so the \nv centers should be closer to the layer than in sample B.
Crystal-TRIM~\cite{posselt1992computer} simulations of the product of implanted nitrogen and created vacancy densities (black line in \autoref{fig:fig1}\,c) yield $45\,\% $ below the cut-off distance.
The product accurately captures the depth distribution of shallow implanted \nv centers, thus serving as a reasonable reference (see Appendix). \\
Because the diamond is exposed to hydrogen plasma before the actual overgrowth begins, these findings suggest that the plasma etches away the outer, possibly damaged diamond layers. 
This would reduce the distances for sample B and increase them for sample A, implying a \nuc layer thinner than the anticipated $1\,\text{nm}$.
Variations in the \nuc concentration can likewise increase the estimated distances.
%

\section{Initialization and control of the nuclear spin layer}
\label{sec:initialization_and_control}
\begin{figure}
    \centering
    \includegraphics[width=0.5\textwidth]{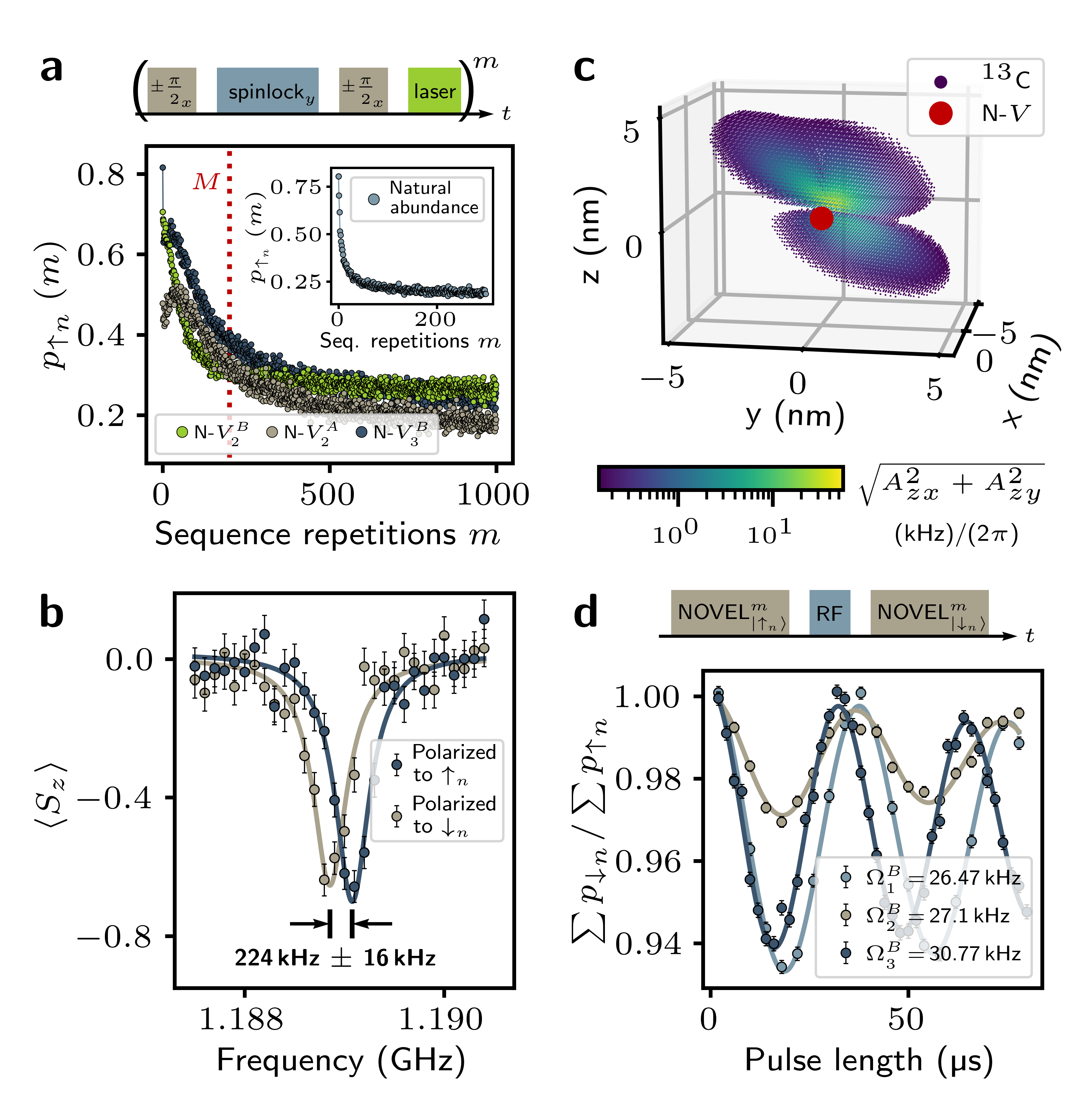}
    \caption{Initialization and control of the nuclear spin layer.
    \textbf{a} NOVEL pulse sequence used to polarize the nuclear spins with typical measurement results.
    The y-axis corresponds to the \nv center's spin flip-flop probability after $m$ sequence repetitions (x-axis) and the inset shows the comparison to a diamond with a natural abundance of \nucn.
    The red vertical line denotes the upper limit $M$ used for the nuclear spin state reconstruction in \autoref{eq:readout}.
    \textbf{b} ODMR measurement of the Overhauser shift (in non-angular frequency units). 
    The color indicates the initialized state of the nuclear spins.
    \textbf{c} Matching the experimental settings, the image visualizes the addressed nuclear spins during the Overhauser shift measurement, assuming a layer thickness of 1\,nm.
    The color of the nuclear spins corresponds to their perpendicular hyperfine coupling, which is proportional to the flip-flop frequency~\cite{london2013detecting}.
    \textbf{d} Rabi measurements of the \nuc layer for different \nv centers, with the pulse sequence.
    After $m$ NOVEL repetitions to polarize the nuclear spins into $\vert{\uparrow_n}\rangle$, an rf pulse is applied, followed by $m$ NOVEL repetitions into $\vert{\downarrow_n}\rangle$ for readout.
    The Rabi frequencies are given in non-angular frequency units.
    The y-axis is proportional to the collective nuclear $\langle I_z\rangle$ expectation value.}
    \label{fig:fig2}
\end{figure}
To initialize the \nuc layer, we employ the nuclear spin orientation via electron spin locking (NOVEL) method~\cite{henstra1988nuclear}, whose pulse sequence is depicted in \autoref{fig:fig2}\,a.
The Rabi frequency of the spinlock pulse is adjusted to match the nuclear Larmor frequency, creating an effective flip-flop interaction via the Hartmann-Hahn condition~\cite{hartmann1962nuclear}.
This allows to pump polarization into the nuclear spin layer through repeated application of the sequence and re-initialization of the \nv center via a laser pulse~\cite{london2013detecting}.
\autoref{fig:fig2}\,a shows typical measurements for up to $m=1000$ sequence repetitions.
The y-axis corresponds to the flip-flop probability $p_{\uparrow_n}(m)$, i.e., the probability to transfer the \nv center's polarization to the nuclear spin layer.
The index $\uparrow_n$ denotes into which state the nuclear spins are initialized, which can be switched to $\downarrow_n$ by reversing the sign of the initial $\pi/2$ pulse.
Compared to natural abundance diamond (inset, \autoref{fig:fig2}\,a), significantly more polarization steps are required to reach a steady state due to the higher number of nuclear spins.
Some measurements show a non exponential decay, which can be attributed to nuclear spins with strong $A_{zx}$ or $A_{zy}$ coupling~\cite{supplement}. 
Such strongly coupled spins can partially inhibit or even fully block the polarization transfer to the other spins~\cite{supplement, whaites2023hyperpolarization, sasaki2024suppression}. \\
Tensor-network simulations show $>90\,\%$ polarization after $2000$ NOVEL repetitions for a planar grid of $25$ spins at the estimated mean distance of $1\,\text{nm}$ to the \nv center~\cite{supplement}.
For more distant nuclear spins, the polarization significantly drops. 
Consequently, the early data points in \autoref{fig:fig2}\,a are dominated by the nearest spins, and the size of the probed spin environment can be adjusted through the total number of repetitions $M$. \\
Polarization of the nuclear spins induces an Overhauser shift~\cite{overhauser1953polarization}, measured in an optically detected magnetic resonance (ODMR) experiment via the \nv center (\autoref{fig:fig2}\,b).
Each data point is obtained after $1000$ NOVEL repetitions and followed by another $1000$ repetitions with an inverted sign of the initial $\pi/2$ pulse to reverse the built-up polarization.  
To ensure that setup instabilities do not affect the measured shift, the measurements for $\vert{\uparrow_n}\rangle$ and $\vert{\downarrow_n}\rangle$ are acquired simultaneously, yielding a shift of $224\,\text{kHz}\pm16\,\text{kHz}$.  
Assuming a $1\,\text{nm}$ thick \nuc layer at the measured distance of $0.8\,\text{nm}$ from the \nv center and neglecting spin diffusion, we estimate an average polarization of $24\,\%\pm2\,\%$ across $\sim15\,000$ nuclear spins from the observed shift.
The estimation only considers spins capable of performing an entire flip-flop within the $1000$ NOVEL repetitions, i.e., within $\sim7\,\text{ms}$ (see \autoref{fig:fig2}\,c).
For $15\,000$ spins the realistic upper bound for 1000 perfect flip-flops is $6.\overline{6}\,\%$ ($=1000/15\,000$), which is much smaller than the polarization estimated from the Overhauser measurement.
This discrepancy is likely caused by nearby spins with strong $A_{zz}$ couplings that enhance the measured sift. \\
With efficient initialization of the nuclear spins in place, we now focus on their control and readout. 
All experiments begin by initializing the nuclear spins into $\vert{\uparrow_n}\rangle$.
Next, the actual experiment is performed, e.g., driving nuclear Rabi oscillations with rf pulses, followed by pumping the spins into $\vert{\downarrow_n}\rangle$. 
For every time step in the Rabi measurements in \autoref{fig:fig2}\,d, we obtain two NOVEL curves, equivalent to those in \autoref{fig:fig2}\,a.
The mean nuclear $\langle I_z\rangle$ expectation value can then be determined from the ratio of summed data points~\cite{supplement}, 
\begin{equation}
    \langle I_z\rangle\propto\sum^M_m p_{\downarrow_n}(m)/ \sum^M_m p_{\uparrow_n}(m).
    \label{eq:readout}
\end{equation}

%

\section{Layer Dynamics and Tunable Interactions}
\begin{figure}
    \centering
    \includegraphics[width=0.5\textwidth]{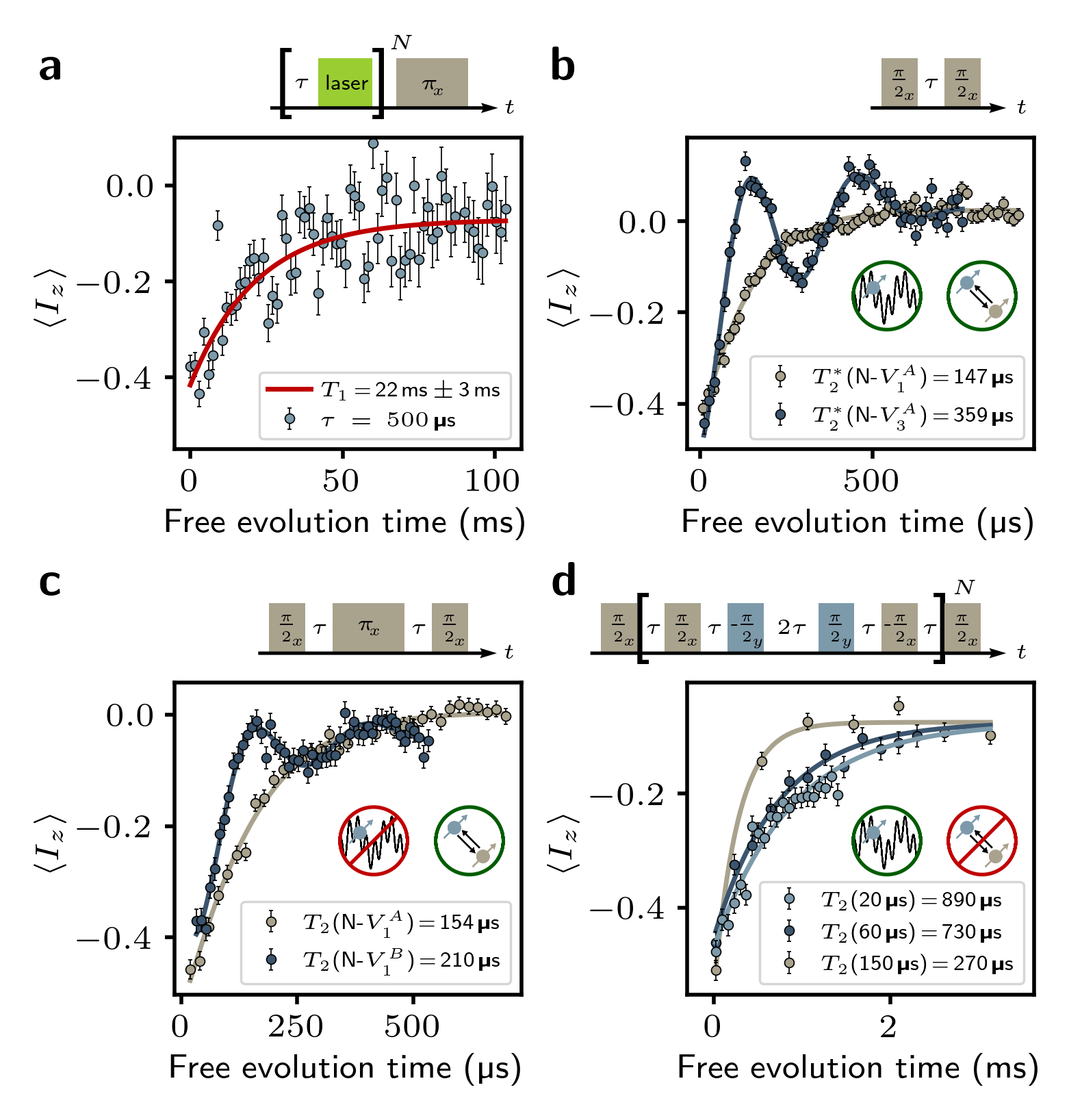}
    \caption{Spin properties of the \nuc layer.
    \textbf{a} Depolarization lifetime measurement in sample B, showing a significantly reduced $T_1$.
    The free evolution time (x-axis) is varied by repeating a free evolution period $\tau$, followed by a laser pulse that resets the \nv center’s spin state to $m_s=0$, for a total of $N$ repetitions.
    \textbf{b} Ramsey measurements with corresponding pulse sequence.
    During the experiment, the nuclear spins are subject to on-site disorder (first pictogram from left) and nuclear dipole-dipole interactions (second pictogram).
    The notation N-$V_k^\mathcal{S}$ indicates the sample $\mathcal{S}$ and the \nv center $k$ used in the measurement. 
    \textbf{c} Designed to suppress on-site disorder, the Hahn echo measurements show minimal improvements and exhibit oscillations attributed to nuclear dipole-dipole interactions.
    \textbf{d} Homonuclear decoupling through the WAHUHA sequence efficiently decouples the nuclear spins from each other.
    The different colors correspond to different pulse spacings $\tau$.} 
    \label{fig:fig3}
\end{figure}

To assess the spin properties of the \nuc layer, we first measure the nuclear depolarization (\autoref{fig:fig3}\,a).  
During the measurement, the \nv center is periodically reset via a laser pulse to account for its limited $T_1$ and maintain the initial $\vert{m_s=0\rangle}$ state. 
The observed depolarization time of $22\,\text{ms}\pm 3\,\text{ms}$ is significantly shorter than the expected $T_1$ at room temperature~\cite{ajoy2019hyperpolarized} and attributed to spin diffusion. 
As polarization spreads and the polarization gradient decreases, diffusion becomes progressively slower, possibly explaining the persistent $z$ magnetization. 
The residual magnetization may also result from normalization errors, e.g., detuning from the Hartmann-Hahn resonance~\cite{supplement}. 
Cycling through the \nv center’s spin states during its reset may also cause additional depolarization. \\
Ramsey measurements reveal nuclear spin dephasing times $T_2^*$ in the low hundreds of microseconds for both samples~\cite{supplement}, as visualized in \autoref{fig:fig3}\,b. 
The observed variation in $T_2^*$ and line shapes likely arises from inhomogeneities of the \nuc concentration.
For comparison, nearly $100\,\%$ \nucn-enriched diamond powder yields $T_2^*=60\,\text{\textmu s}\pm 5\,\text{\textmu s}$ at $7.05\,\text{T}$ in an NMR measurement~\cite{supplement}.
Apart from inhomogeneities, the longer $T_2^*$ of our \nuc layer can be attributed to its thickness of only a few atomic layers, which limits nuclear interactions. \\ 
Measuring the nuclear coherence time via a Hahn echo~\cite{hahn1950spin} leads to only minor improvements over $T_2^*$.
As seen in \autoref{fig:fig3}\,c, oscillations arising from dipolar interactions among nuclear spins persist, since spin echo sequences like Hahn echo are designed to decouple on-site disorder~\cite{choi2020robust}, e.g., detuning. 
Assuming dephasing is dominated by dipolar interactions, a Hahn echo is not expected to extend the coherence time. \\ 
In contrast, homonuclear decoupling sequences like WAHUHA~\cite{waugh1968approach, choi2020robust}, which decouple dipolar interactions, significantly increase the coherence time, as illustrated by \autoref{fig:fig3}\,d. 
Effective decoupling requires pulse spacings $\tau$ much shorter than $T_2^*$. 
If $\tau$ approaches $T_2^*$, the decoupling fails. 
The WAHUHA sequence not only confirms strong interactions among the nuclear spins but also allows to tune their interaction strength by varying $\tau$.
At short $\tau$, our measurements are limited by rf heating due to the increased duty cycle. 
%

\section{Creation of Discrete Time-Crystalline Order}
\begin{figure*}
    \centering
    \includegraphics[width=1\textwidth]{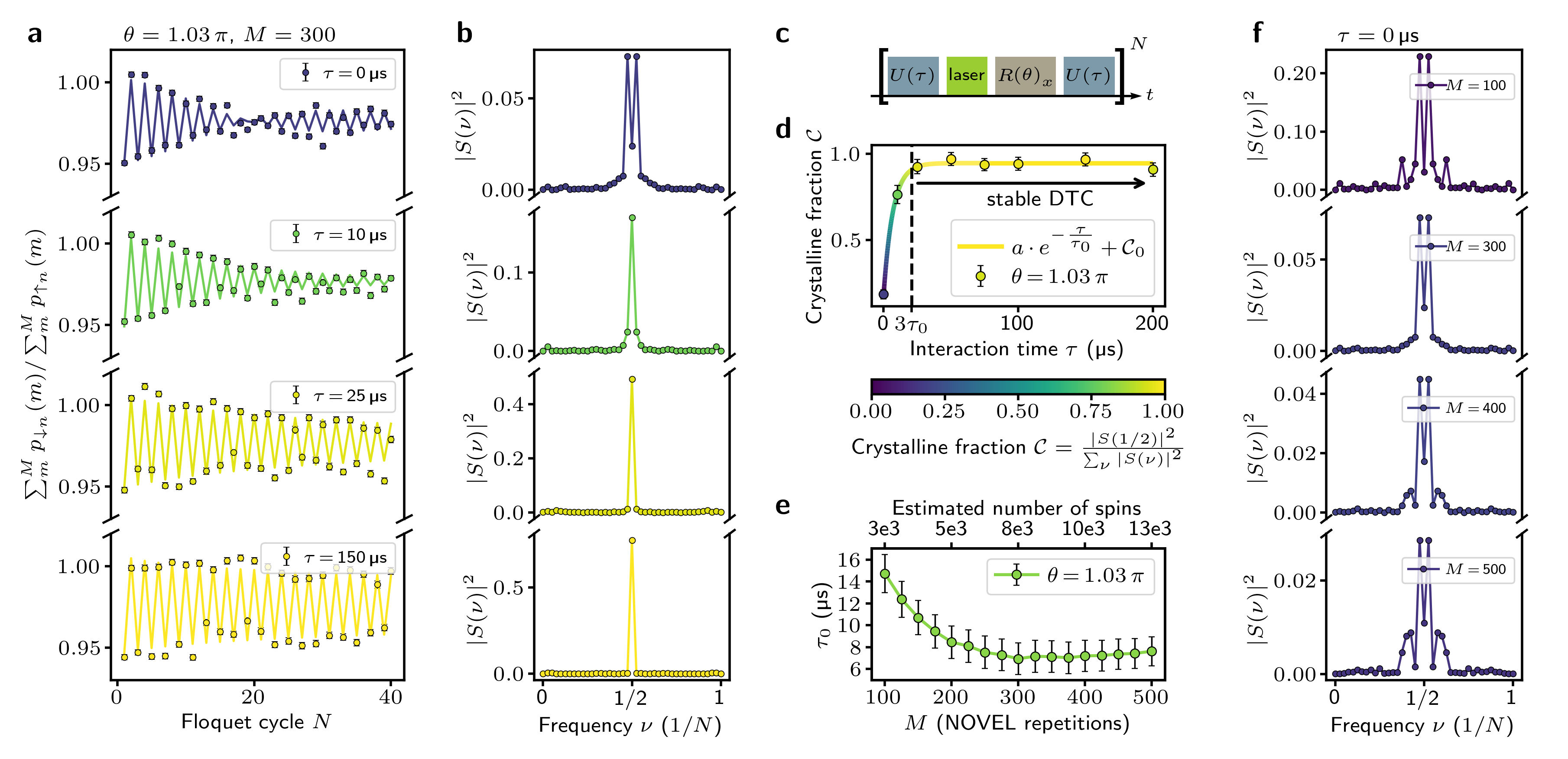}
    \caption{Exploring discrete time-crystalline order with the nuclear spin layer.
    \textbf{a} Representative discrete time-crystal (DTC) measurements for different interaction times $\tau$.
    For $\tau=0\,\text{\textmu s}$, the signal shows a beating caused by the applied over-rotation with rotation angle $\theta=1.03\,\pi$, which vanishes for increasing $\tau$.
    The x-axis corresponds to the Floquet cycles $N$ and the solid lines show the fit result.
    We use $a \cdot \cos(\pi p\cdot N\tau )\cdot \exp(-(N\tau/T)^\beta) + c$ as the fit function, with $a,p,T,\beta$ and $c$ as free parameters.
    The y-axis shows the summed, normalized NOVEL signal ($\propto\langle I_z\rangle$), see \autoref{eq:readout}.
    \textbf{b} Power spectral density of the measurement data for different interaction times $\tau$. 
    Each row corresponds to the data presented in panel a. 
    \textbf{c} Pulse sequence of the DTC experiment.
    The blue blocks correspond to a free evolution time for duration $\tau$ and the gray block denotes a rf rotation pulse of angle $\theta$. 
    The \nv center is re-initialized into $m_s=0$ through a laser pulse in each repetition $N$.
    \textbf{d} Calculating the crystalline fraction $\mathcal{C}$ for different interaction times $\tau$ allows to visualize the transition to the DTC phase.
    The measurement data are fitted with a single exponential characterized by a $1/e$ time $\tau_0$. 
    \textbf{e} Extracted values of $\tau_0$ for increasing number of NOVEL repetitions $M$. 
    The second x-axis shows the estimated number of nuclear spins capable of performing a full flip-flop within $M$ NOVEL repetitions.
    For low $M$, where spins near the ${}^{12}\text{C}$-${}^{13}\text{C}$ interface dominate the signal, a stable DTC phase is reached more slowly than for larger $M$, likely due to reduced average spin–spin coupling in the more dilute interface region.
    \textbf{f} Power spectral density for $\tau=0\,\text{\textmu s}$ at different values of $M$.
    Varying $M$ probes different dynamics within the layer, reflected in additional beatings arising from the induced over-rotation.
    }
    \label{fig:fig4}
\end{figure*}
To demonstrate the \nuc layer's applicability to explore many-body interactions, we use it to probe time-crystalline order.
A time crystal is a non-equilibrium phase of matter that spontaneously breaks time-translation symmetry~\cite{wilczek2012quantum, zaletel2023colloquium}. 
It can occur in many quantum systems, most notably periodically driven, so-called Floquet systems~\cite{choi2017observation, randall2021many, mi2022time}, where it manifests as a discrete time crystal (DTC), exhibiting oscillations at multiples of the driving period that persist under perturbations~\cite{yao2017discrete}. \\
To create a DTC with our \nuc layer, we apply the pulse sequence shown in \autoref{fig:fig4}\,c.
Each Floquet cycle $N$ ($\equiv$ one sequence repetition) comprises a rotation pulse $R(\theta)$ with angle $\theta$ placed between two free evolution times $U(\tau)$ of length $\tau$. 
After each $U(\tau)$, a laser pulse resets the \nv center to $m_s=0$. 
Measurements for sample B are shown in \autoref{fig:fig4}\,a, choosing $\theta=1.03\,\pi$ for varying $\tau$.
In a non-interacting system, the 3\,\% over-rotation creates a beating that we observe for $\tau=0\,\text{\textmu s}$, accompanied by a strong decay.
\autoref{fig:fig4}\,b shows the corresponding power spectral density of the baseline-corrected measurement data.
The baseline itself is determined from the fits shown as solid lines in \autoref{fig:fig4}\,a.
Increasing $\tau$ allows the nuclear spins to interact, forming the time-crystalline phase that stabilizes the oscillation with the expected 2$T$-periodic response, where $T$ is the Floquet period. \\
By calculating the crystalline fraction $\mathcal{C}=|S(\nu=1/2)|^2/\sum_\nu|S(\nu)|^2$~\cite{choi2017observation}, the ratio of the $\nu=1/2$ amplitude (intensity of the $2T$-periodic response) to the total spectral power, we can visualize the phase transition in \autoref{fig:fig4}\,d. 
In contrast to experiments utilizing layers of coupled \nv centers~\cite{choi2017observation, wu2025spin, gao2025signal}, our system allows us to bias the experiment toward different spatial regions and coupling classes of the layer during post-processing by tuning $M$ (the number of NOVEL repetitions; see \autoref{eq:readout}).
As discussed in \autoref{sec:initialization_and_control}, for small $M$ the dynamics are dominated by the nearest nuclear spins to the \nv center, whereas more distant spins become dominant at larger $M$.
By phenomenologically approximating the observed phase transition with a single exponential (with $R^2>95\,\%$ for all curves), we can extract the $1/e$ time $\tau_0$ for different values of $M$, to visualize the differences in the observed dynamics in \autoref{fig:fig4}\,e.
The second x-axis of \autoref{fig:fig4}\,e shows the estimated number of nuclear spins read out, assuming a layer thickness of $1\,\text{nm}$ and restricting the interaction range to spins capable of completing a full flip-flop within $M$ NOVEL repetitions.
For small $M$, i.e., when the signal is dominated by the nearest nuclear spins, we observe a significantly longer interaction time required for the system to exhibit stable DTC order ($\tau>3\tau_0$) compared to larger $M$.
This behavior is likely caused by the reduced nuclear–nuclear coupling strength of the nearest spins, which are located at the ${}^{12}{\text{C}}$-${}^{13}{\text{C}}$ interface. 
In contrast, more distant spins, situated deeper within the \nuc layer, experience stronger average coupling and can interact with a larger number of neighboring nuclear spins, leading to a much shorter $\tau_0$ at large values of $M$.
This effect is also evident in \autoref{fig:fig4}\,f at an interaction time of $\tau=0\,\text{\textmu s}$, where no DTC order has yet emerged. 
For $M=300$, the dynamics differ significantly from those observed at $M=500$, where variations in coupling strength give rise to additional beatings caused by induced over-rotation ($\theta=1.03\,\pi$).
At large $M$, this also results in a slight increase in $\tau_0$ (within the margin of error), which may indicate the presence of inhomogeneities or patchiness in the \nuc layer. \\
Owing to the similarity to prior experiments with dense \nv center ensembles~\cite{choi2017observation}, specifically the long-range interaction among the spins, our time-crystalline order is likely stabilized by slow, critical dynamics~\cite{ho2017critical} rather than many-body localization~\cite{randall2021many}. 
Because the decay rate of the DTC order does not increase with interaction time~\cite{supplement}, prethermalization~\cite{beatrez2021floquet, peng2021floquet} can also be ruled out as a stabilization mechanism. \\
Our DTC measurements not only serve as a proof-of-concept demonstration for probing many-body interactions in a large interacting nuclear spin layer, but also highlight the modular readout capabilities of our hybrid \nvn-\nuc system, enabling the investigation of different dynamics and settings within a single measurement.

\section{Conclusion}

In conclusion, we present a room-temperature platform to probe many-body phenomena based on an atomically thin nuclear spin layer, along with its fabrication, characterization, and operation.   
The control could be extended by combining our system with scanning probe techniques, such as magnetic tips~\cite{mamin2012high}, to induce strong field gradients that locally modify the layer's Hamiltonian. 
With an estimated layer thickness below $1\,\text{nm}$, regions with two-dimensional characteristics are within reach and may already contribute to the observations. 
Such low-dimensional, driven systems are expected to change the relaxation dynamics~\cite{burin2015localization, ho2017critical}, facilitating the investigation of novel phenomena. 
For example, in a low, tilted magnetic field or in a [111] diamond, all nuclear spins are oriented perpendicular to the two-dimensional layer, enabling the generation of spin squeezing~\cite{wu2025spin}.
Apart from varying the \nv center's sensing volume, changing the \nuc concentration also allows to control the disorder, a key parameter when studying time-crystalline order~\cite{yao2017discrete, beatrez2021floquet}. 
Dynamical decoupling schemes like the showcased AXY sequence~\cite{casanova2017arbitrary} or the dynamically decoupled radio-frequency (DDrf) sequence~\cite{bradley2019ten-qubit} permit selective control \nuc nuclear spins via the \nv center, rendering the system partially programmable. \\ 
Given the simplicity and accessibility of our system, our work provides a convenient and scalable platform to study many-body quantum phenomena at room temperature.
%

%
%


\section{Acknowledgments}   

We thank Ashok Ajoy, Martin Korzeczek, Michael Olney-Fraser, Priyadharshini Balasubramanian, and Thomas Reisser for helpful discussions. 
We thank Matthias Posselt and the Helmholtz-Zentrum Dresden-Rossendorf for providing the Crystal-TRIM software. 
J.P. acknowledges support from the QuantERA II Programme (Mf-QDS) that has received funding from the European Union’s Horizon 2020 research and innovation program under Grants No. 101017733, from Grants No. TED2021-130578BI00 and grant PID2021-124965NB-C21 funded by MICIU/AEI/10.13039/501100011033 and by the European Union NextGenerationEU/PRTR and from Quantum Flagship project ASPECTS, Grant Agreement No. 101080167.
This work was funded by the German Federal Ministry of Research (BMBF) by future cluster QSENS (No. 03ZK110AB) and projects DE-Brill (No. 13N16207), SPINNING (No. 13N16215), PhoQuant (No. 13N16110), DIAQNOS (No. 13N16463), quNV2.0 (No. 13N16707), QR.N (No. 16KIS2208) and EXTRASENS (No. 13N16935), DLR via project QUASIMODO (No. 50WM2170), Deutsche Forschungsgemeinschaft (DFG) via Projects No. 386028944, No. 387073854, No. 445243414, No. 491245864, No. 499424854, No. 532771161, No. 546850640, and joint DFG/JST ASPIRE program via project  No. 554644981, European Union's HORIZON Europe program via projects QuMicro (No. 101046911), SPINUS (No. 101135699), C-QuENS (No. 101135359), QCIRCLE (No. 101059999), and FLORIN (No. 101086142), European Research Council (ERC) via Synergy grant HyperQ (No. 856432), and Carl-Zeiss-Stiftung.

\section{Author Contributions}

P.J.V., M.B.P., J.P. and F.J. conceived the experiments.
P.J.V. conducted and analyzed the measurements of the \nuc layer samples and performed the simulations for the time-crystalline order.
C.F., C.O. and J.L. fabricated the samples.
R.B. performed the NMR measurements and C.F. measured the depths of the 5\,keV implanted \nv centers.
The Crystal-TRIM simulations were carried out by J.F..
M.K. developed the numerical method and model for the distance estimation and produced numerical data which P.J.V. used to determine the distance from experimental data.
A.V. and J.P. performed the tensor network simulations for the polarization dynamics.
J.P., M.B.P. and F.J. supervised the work.
All authors discussed the results and contributed to the manuscript.

\section{Data Availability}

All data presented in this study are available from the corresponding authors on reasonable request.

\section{Competing Interests}

C.F., C.O., J.L. and F.J. are co-founders, C.F. is CTO, C.O. is CFO \& COO and J.L. is CEO of Diatope GmbH.
P.J.V. was employed part-time by Diatope GmbH during parts of the project.


\appendix

\section{Overgrowth Settings}

In-between all fabrication steps that require the diamonds to leave or re-enter the CVD reactor, the samples are cleaned by a 1:1:1 mixture of nitric ($65~\%$), sulfuric ($98~\%$) and perchloric ($72~\%$) acid for 20 minutes at 200°C in a microwave digestion system (ETHOS.lab).
After the acid cleaning, they are rinsed with deionized water. \\
The growth process begins by heating the reactor's sample plate to 700\,${}^{\circ}$C with a graphite substrate heater.
We subject the diamonds to a hydrogen plasma for 5\,min to ensure the plasma and temperature are in equilibrium before we start the overgrowth.
The temperature is monitored by an infra-red pyrometer (Optris, model CTlaser 2 MH, wavelength 1.6\,\textmu m).
During the overgrowth we reach a temperature of approximately 900\,${}^{\circ}$C.
We use isotope-enriched methane from Cambridge Isotope Laboratories with a purity of 99.999\,\% for \ctw and 99.9\,\% for \nucn.
The gases are additionally purified through a palladium filter (Johnson Matthey, Hydrogen Purifier HP-25) and a getter filter for methane (MonoTorr, PS4-MT3-531). 
\autoref{tab:methods_tab1} summarizes the growth parameters for each overgrowth step.
The growth rates are estimated from our earlier work~\cite{findler2020indirect}.
The diamond surface corresponds to the [100] plane, with sample A exhibiting a miscut angle of 0.3${}^\circ$ and sample B 0.1${}^\circ$, as determined by X-ray diffraction (XRD) rocking curve analysis.
A comprehensive description of the CVD reactor can be found in~\cite{silva2010microwave, osterkamp2021engineering}. \\
The implantation is performed in a home-built ion implanter~\cite{lang2023atomic}, that features an IQE12/38 ion source from Specs.
To distinguish implanted \nv centers from native ones, we choose a 98\,\% enriched ${}^{15}\text{N}_2$ gas from Sigma-Aldrich (${}^{15}\text{N}$ natural abundance: 0.4\,\%), which is purified by a Wien mass filter during the implantation to ensure that we only implant ${}^{15}\text{N}^+$.
Using an implantation energy of 1\,keV, we create several implantation spots with a diameter of 200\,\textmu m.
We choose three different doses of $1\cdot10^{11}\frac{{}^{15}\text{N}^+}{\text{cm}^2}$, $2\cdot10^{11}\frac{{}^{15}\text{N}^+}{\text{cm}^2}$, and $3\cdot10^{11}\frac{{}^{15}\text{N}^+}{\text{cm}^2}$ to create single \nv centers at varying densities after the overgrowth~\cite{findler2020indirect}. \\ 
To finally form the \nv centers from the implanted nitrogen, we place the samples in a home-built ultra-high-vacuum oven at $1000\,{}^{\circ}\text{C}$ for $3$ hours with a pressure below $1\cdot10^{-7}\,\text{mbar}$ during the annealing process. \\
Before their investigation in a room-temperature confocal microscope, the diamonds are acid cleaned one more time following the cleaning procedure described at the beginning of this Appendix. \\
\begin{table}
\centering
\caption{Growth parameters for the different overgrowth steps. The columns show the current step of the overgrowth process and the rows show the corresponding growth parameters.}
\renewcommand{\arraystretch}{1.5}
\begin{tabular}{| c | >{\centering\arraybackslash}p{1.45cm} | >{\centering\arraybackslash}p{1.45cm} | >{\centering\arraybackslash}p{1.45cm} |}
  \hline
  \diagbox{\textbf{Parameter}}{\textbf{Step}} & \textbf{1} & \textbf{2} & \textbf{3} \\  
  \hline
  \textbf{Methane Gas} & ${}^{12}\text{CH}_4$ & ${}^{13}\text{CH}_4$ & ${}^{12}\text{CH}_4$ \\
  \hline
  \textbf{Purity} & 99.999\,\% & 99.9\,\% & 99.999\,\% \\
  \hline
  \textbf{$\text{CH}_4$:$\text{H}_2$} & 0.083\,\% & 0.025\,\% & 0.083\,\% \\
  \hline
  \textbf{Flow Rate $\text{H}_2$} & \multicolumn{3}{|c|}{600\,sccm} \\
  \hline
  \textbf{Pressure}  & \multicolumn{3}{|c|}{22.5\,mbar} \\
  \hline
  \textbf{Microwave Power} & \multicolumn{3}{|c|}{1.2\,kW} \\
  \hline
  \textbf{Temperature} & \multicolumn{3}{|c|}{900\,${}^{\circ}$C} \\
  \hline
  \textbf{Duration} & 180\,min & 11\,min & 60\,min \\
  \hline
  \hline
  \textbf{Target Thickness} & 96\,nm & 1\,nm & 32\,nm \\
  \hline
\end{tabular}
\label{tab:methods_tab1}
\end{table}

\section{Distance Estimation}

\begin{figure}
    \centering
    \includegraphics[width=0.5\textwidth]{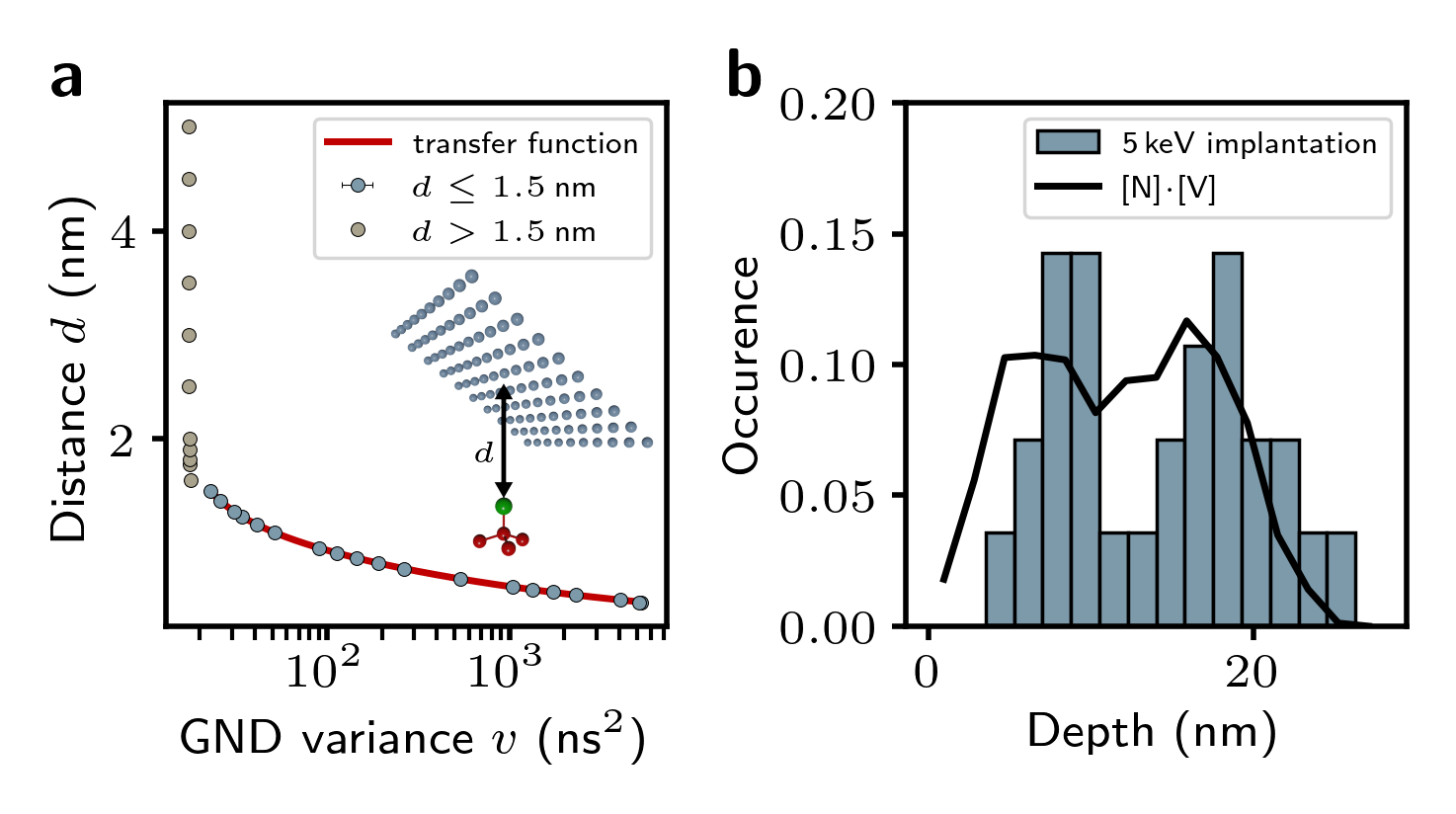}
    \caption{Distance estimation. 
    \textbf{a} Corresponding distances of the extracted variances of the fitted, simulated AXY spectra. 
    The red line (fit) shows the transfer function that is used to estimate the distance of the \nv center to the \nuc layer.
    \textbf{b} Depth distribution of a 5\,keV implantation.
    The black solid line shows the product of the simulated vacancy density [V] and nitrogen density [N].}
    \label{fig:methods_fig2}
\end{figure}

To infer the distance of the \nv centers to the \nuc layer from the AXY spectra, we start by numerically simulating the experiment.
Equivalent to the experiment we use $30$ repetitions of the AXY$8$ sequence with the Fourier coefficients $\{f_1=0.1,f_2=f_3=f_4=0\}$~\cite{casanova2015robust}.
The \nuc layer is modeled by a two-dimensional rectangular grid of 100 \nuc nuclear spins, placed at distances of 0.154\,nm, corresponding to the next-neighbor distance in the diamond lattice. 
Our model captures two-spin correlations among the nuclear spins as well as three-spin correlations mediated by the \nv center.
A detailed description of the model is found in the Supplemental Material~\cite{supplement}. \\
The simulated spectra are fitted with a generalized normal distribution~\cite{nadarajah2005generalized} in the time-domain (pulse spacing $\tau=1/(2\nu)$) and the extracted variances $v$ for various distances are shown in \autoref{fig:methods_fig2}\,a.
Beyond 1.5\,nm the signal's amplitude vanishes and the linewidth is solely given by the filter created by the AXY sequence.
Therefore, we limit our estimation to distances up to $1.5\,\text{nm}$ and phenomenologically fit the obtained variances as a function of distance with the transfer function $d(v)=a(v-v_0)^{-b}+d_0$.
We obtain $a=2.222\,\frac{\text{nm}}{\text{ns}^{-2b}} \pm 0.017\,\frac{\text{nm}}{\text{ns}^{-2b}}$, $b=0.221 \pm 0.005$, $v_0=14.87\,\text{ns}^2 \pm 0.25\,\text{ns}^2$, and $d_0=0.0950\,\text{nm} \pm 0.0127\,\text{nm}$, with an $R^2$ of $99.995\,\%$. \\
Applying the same fit routine to the experimental data allows us to then estimate the \nv center-\nuc layer distance from the extracted variance via the transfer function.
Depending on the number of observed resonances, the data are fitted with either a single or double generalized normal distribution, combined with an exponential decay to account for the limited coherence time. 
Data points that are attributed to sharp nitrogen resonances from imperfect magnetic field alignment are omitted~\cite{joas2025high}. \\ 
All extracted variances and estimated distances can be found in the Supplemental Material~\cite{supplement}.

\section{5\,keV Implantation Comparison}
\autoref{fig:methods_fig2}\,b shows the depth distribution of a $5\,\text{keV}$ ${}^{15}\text{N}^+$ implantation into a \ctw overgrown diamond.
The depths are measured for $28$ \nv centers via the XY8 sequence~\cite{pham2016nmr}.
We simulate the implantation with Crystal-TRIM~\cite{posselt1992computer} and calculate the product of the vacancy and nitrogen densities as comparison, which is shown by the black solid line in \autoref{fig:methods_fig2}\,b.
The simulation accurately reflects the profile of the observed distribution but is shifted to smaller depths.
This shift can be explained by a slightly different implantation angle, which can alter the channeling effects and thus the final depths.
Nevertheless, the simulated product provides a valuable reference for the expected \nv center-\nuc layer distance distribution in our samples.

\section{Error Estimation}

The \nv center's normalized fluorescence is obtained by dividing the sum of photon counts during the first $\sim350\,\text{ns}$ by that of the steady-state (recorded for $1.6\,\text{\textmu s}$) under green ($\lambda=561\,\text{nm}$) illumination.
The measurement error is derived from the corresponding shot noise~\cite{degen2017quantum} and propagated using Gaussian error propagation.
Unless stated otherwise, all errors reflect the propagated errors of the original fluorescence data.
For fit results or quantities derived from fit results, the error is taken as the standard deviation of the fit parameter and propagated accordingly.


\bibliography{references}

%
%
\end{document}


\title{Supplementary Information for: Probing Many-Body Phenomena with Atomically Thin Nuclear Spin Layers in Diamond}

\author{Philipp J. Vetter}
\email{philipp.vetter(at)uni-ulm.de}
\affiliation{Institute for Quantum Optics, Ulm University, Albert-Einstein-Allee 11, 89081 Ulm, Germany}
\affiliation{Center for Integrated Quantum Science and Technology (IQST), 89081 Ulm, Germany}
\author{Christoph Findler}
\affiliation{Institute for Quantum Optics, Ulm University, Albert-Einstein-Allee 11, 89081 Ulm, Germany}
\affiliation{Center for Integrated Quantum Science and Technology (IQST), 89081 Ulm, Germany}
\affiliation{Diatope GmbH, Buchenweg 23, 88444 Ummendorf, Germany}
\author{Antonio Verdú}
\affiliation{Departamento de Física, Universidad de Murcia, 30071 Murcia, Spain}
\author{Matthias Kost}
\affiliation{Institute of Theoretical Physics, Ulm University, Albert-Einstein-Allee 11, 89081 Ulm, Germany}
\affiliation{Center for Integrated Quantum Science and Technology (IQST), 89081 Ulm, Germany}
\author{Rémi Blinder}
\affiliation{Institute for Quantum Optics, Ulm University, Albert-Einstein-Allee 11, 89081 Ulm, Germany}
\affiliation{Center for Integrated Quantum Science and Technology (IQST), 89081 Ulm, Germany}
\author{Jens Fuhrmann}
\affiliation{Institute for Quantum Optics, Ulm University, Albert-Einstein-Allee 11, 89081 Ulm, Germany}
\affiliation{Center for Integrated Quantum Science and Technology (IQST), 89081 Ulm, Germany}
\author{Christian Osterkamp}
\affiliation{Diatope GmbH, Buchenweg 23, 88444 Ummendorf, Germany}
\author{Johannes Lang}
\affiliation{Diatope GmbH, Buchenweg 23, 88444 Ummendorf, Germany}
\author{Martin B. Plenio}
\affiliation{Institute of Theoretical Physics, Ulm University, Albert-Einstein-Allee 11, 89081 Ulm, Germany}
\affiliation{Center for Integrated Quantum Science and Technology (IQST), 89081 Ulm, Germany}
\author{Javier Prior}
\affiliation{Departamento de Física, Universidad de Murcia, 30071 Murcia, Spain}
\author{Fedor Jelezko}
\email{fedor.jelezko(at)uni-ulm.de}
\affiliation{Institute for Quantum Optics, Ulm University, Albert-Einstein-Allee 11, 89081 Ulm, Germany}
\affiliation{Center for Integrated Quantum Science and Technology (IQST), 89081 Ulm, Germany}
%


\maketitle

\tableofcontents

\section{Spin Properties of the \nv \,Centers}

\begin{figure}[h]
    \centering
    \includegraphics[width=0.48\textwidth]{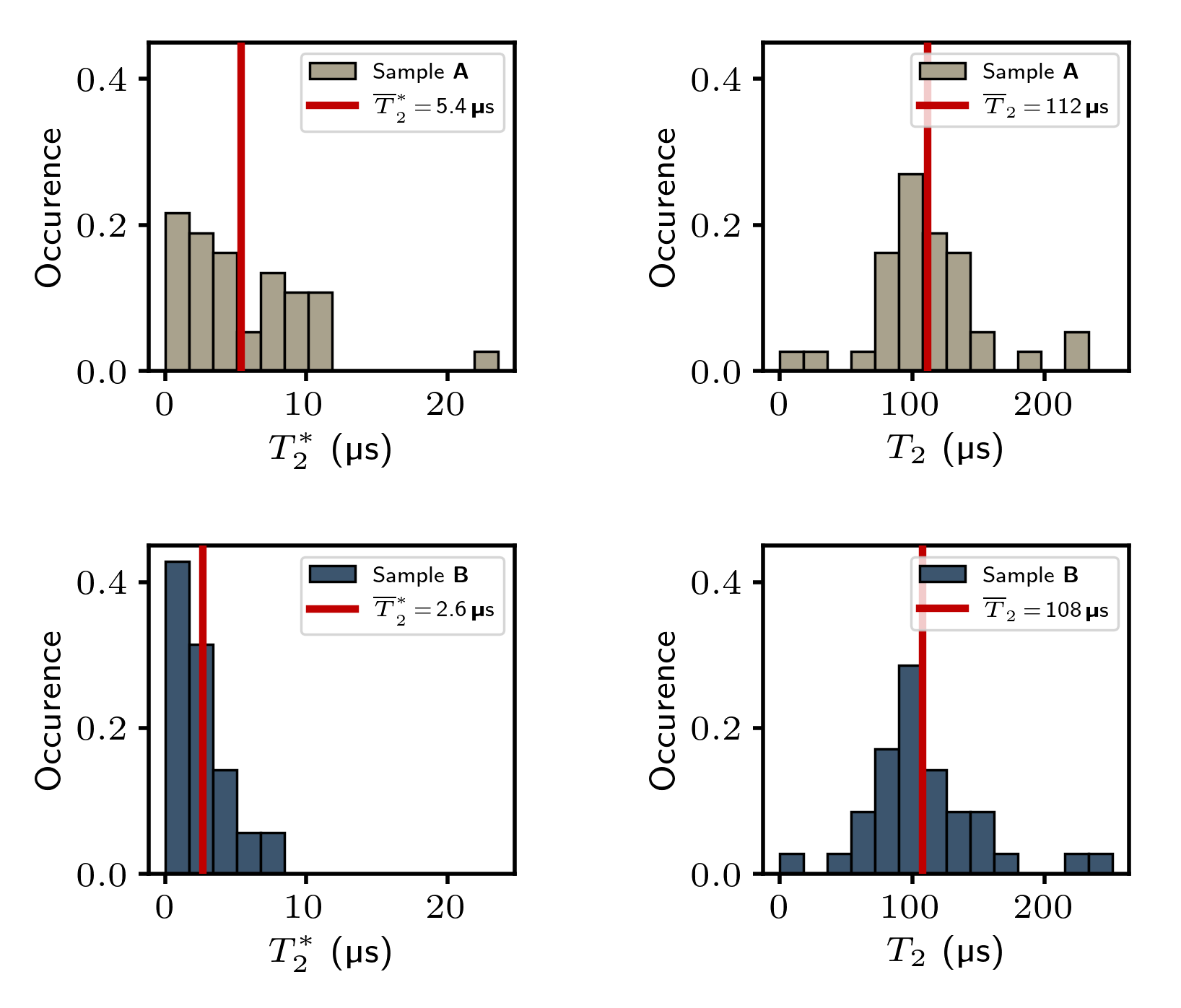}
    \caption{Dephasing and coherence times of the overgrown \nv centers.
    \textbf{a} Dephasing time $T_2^*$ distribution and \textbf{b} Hahn Echo coherence time $T_2$ distribution for sample A.
    The red line shows the corresponding mean value.
    \textbf{c} $T_2^*$ and \textbf{d} $T_2$ distribution for sample $B$.
    All data sets consist of more than $35$ measurements.}
    \label{fig:supplementary_fig_1}
\end{figure}

To demonstrate the feasibility of our overgrowth method, we investigate the coherence properties of the implanted \nv centers.
For each sample we perform more than 35 Ramsey and Hahn echo~\cite{hahn1950spin} experiments leading to an average dephasing time of $\overline{T}_{2,A}^*=5.4\,\text{\textmu s}\pm0.8\,\text{\textmu s}$ and a coherence time of $\overline{T}_{2,A}=112\,\text{\textmu s}\pm 7\,\text{\textmu s}$ for sample A and $\overline{T}_{2,B}^*=2.6\,\text{\textmu s}\pm0.4\,\text{\textmu s}$ and $\overline{T}_{2,B}=108\,\text{\textmu s}\pm8\,\text{\textmu s}$ for sample B.
The reported errors denote the uncertainty of the mean.
\autoref{fig:supplementary_fig_1} visualizes the corresponding distributions as histograms.
The average coherence times are comparable to \nv centers located at similar depths in pure \ctw diamond~\cite{findler2020indirect}.
The dephasing times are smaller than those of pure \ctw diamond, likely due to strong coupling to nearby nuclear spins.
In addition, we observe a large variance of $T_2^*$ and $T_2$ due to the varying distances to the \nuc layer. \\
We observe slightly longer dephasing and coherence times in sample A than in sample B.
The prolonged $T_2^*$ and $T_2$ indicate a larger average distance to the \nuc layer, assuming the \nuc nuclear spins as the main source of decoherence. 
While not performed in the same region, the results match the observations of the AXY distance measurements presented in the main text. \\
A more detailed analysis using explicit noise models for \nuc layers of varying thickness, density, and distance could yield further insights.

\section{AXY Spectra Simulation}

To address the large number of spins required to capture the spectral characteristics of a thin spin layer, we distribute 100 $^{13}\mathrm C$ nuclear spins on a rectangular two-dimensional grid with a nearest-neighbor distance of $0.154\,\text{nm}$.
The layer is tilted by 54.7${}^{\circ}$ to the symmetry axis of the \nv center to account for the [100] growth direction.
All interactions are modeled as dipole-dipole interactions, which, in the case of the \nv center-\nuc coupling is valid for $\vert\vec{r}\vert \gtrsim 2a_0$~\cite{gali2008ab-into, blinder2024reducing}.
$a_0=0.357\,\text{nm}$ is the diamond lattice constant at room-temperature~\cite{hom1975accurate}.
Below this threshold, the \nv center's spin density cannot be neglected and can lead to stronger hyperfine couplings.
Due to the \nv center's large zero-field splitting, we apply the secular approximation~\cite{maze2012free}, which prohibits flips of the \nv center, removing all non-$S_z$-coupling terms. \\
The description of the multi-spin state covers all polarizations and multi-spin correlations up to a fixed length in a large array of coefficients.
To achieve a high level of compression, we discard any terms that require more than two non-identity Pauli operators within the subset of \nuc spin bases.
We compute in advance how, within such a state description, polarizations and correlations interact with coefficients from the Hamiltonian's terms, given the structure of the Hamiltonian and the effect of pulses within the AXY sequence.
The actions require to perform a single order timestep via Euler-forward iteration, defined by the action of the Hamiltonian commuting with a general state. 
We precompute and store this action on the level of index operations acting on the input state, the Hamiltonian's coefficients, and the target state array.
When we compute the dynamics, we repeatedly apply the set of these index operations to the state, to achieve higher orders in the time evolution, i.e.
\begin{equation}
\frac{\mathrm d}{\mathrm dt}\rho^{(k)} = \rho^{(k+1)} = i \left[\rho^{(k)}, H\right]
\end{equation}
and the whole timestep expands to
\begin{equation}
\begin{split}
&\rho(t+\Delta t) - \rho(t) \\
&= \left(\Delta t\right) \rho^{(1)}  + \frac{\left(\Delta t\right)^2}2 \rho^{(2)} + \frac{\left(\Delta t\right)^3}6 \rho^{(3)} + \dots \;.
\end{split}
\end{equation}
Therefore, the action of a Hamiltonian on a state, both represented by the coefficients of a previously fixed set of terms, is completely defined by the list of entry operations between the two representing arrays that lead to non-zero and non-truncated contributions in the target array. \\
The initial state of the simulation has all spins except the \nv center spin completely unpolarized.
The rate at which correlations can build up is limited by the strength of the interaction between the spins.
Within the distances considered in the simulations, the \nv center is sufficiently close to the spin layer so that the coupling strength between its electron spin and the \nuc spins dominates the polarization dynamics over the interaction between the \nuc spins within the layer.
To account for this strong interaction, three-spin-correlations are considered when they include the \nv center as one of their constituents while discarded otherwise.
Consequently, all two-spin correlations are included, while correlation terms of order four or higher are discarded.
Starting from an uncorrelated state and having only weak build-up paths for correlations between the \nuc spins via their weak direct interaction or a third spin, we can capture the relevant polarization dynamics that influence the shape and characteristics of the resulting AXY spectra. \\
The pulse sequence is simulated via perfect pulses, i.e. infinitely short and completely lossless.
We combine the pulse applications with the individual free evolution times in-between the pulses according to the frequency matching condition in the AXY sequence and finally read the \nv center polarization at the end of the sequence from its $z$-polarization entry in the state vector to get the numerical value of the spectrum at the measured frequency.
Although the free evolution times between the pulses vary for different resonance frequencies, this procedure can be highly parallelized.
Instead of varying the amount and length of time steps, we rescale the magnitude of the Hamiltonian terms, according to
\begin{equation}
f(\alpha t,\, \hat H) = f(\alpha t \hat H) = f(t, \alpha \hat H)
\end{equation}
as the time parameter $t$ and the Hamiltonian operator $\hat H$ always show up as a product in the equations that govern the dynamics. \\
Our implementation requires approximately 20 hours on an NVIDIA A100 GPU for a run of 50 AXY-8 cycles with 100 \nuc spins, evaluating 200 frequencies in parallel.
The performance of the implementation is limited by the available memory bandwidth, 
as the time used for the large number of operations to be performed for a timestep is mainly used by the memory lookup of indices and the subsequent lookup of their corresponding values, which is significantly slower than the simple floating point multiplication that follows. 
The truncation of the subspace by correlation length leaves us with a polynomial scaling in complexity and enables us to study the dynamics of spin layers of arbitrary geometries. \\
To validate our simulation results, we varied the setup parameters and studied their impact on the resulting spectra. 
We have numerically verified that the weaker coupled spins close to the border of the simulated layer do not alter the width of the main spectral peak, as the \nv electron spin gradient locates their influence close to the center of the spectrum. 
Both the consideration of two-spin correlations among the \nuc spins and three-spin correlations via the \nv center impact the width of the \nuc resonance, which we tested by suppressing them individually.
To test the effect of higher-order correlations, we included three-spin-correlations within the \nuc spins.
We kept them within geometrical proximity to bound the growth in numerical complexity, but found no effect that would exceed minor impact on the individual shape of the central part of the resulting spectrum, especially no effect on its width.
Finally, we compared a numerically exact simulation of 8 spins against the compressed simulation approach, where we found high consistency of the spectra even for long timescales and stronger interactions. 
Although the qualitative details of the spectrum showed tolerable deviations after sufficiently many iterations of the AXY sequence, the qualitative details, such as width and coarse shape of the peak, remained consistent within the whole comparison.

\subsection{Comparison with Multiple Layers}

\begin{figure}[h]
    \centering
    \includegraphics[width=0.5\textwidth]{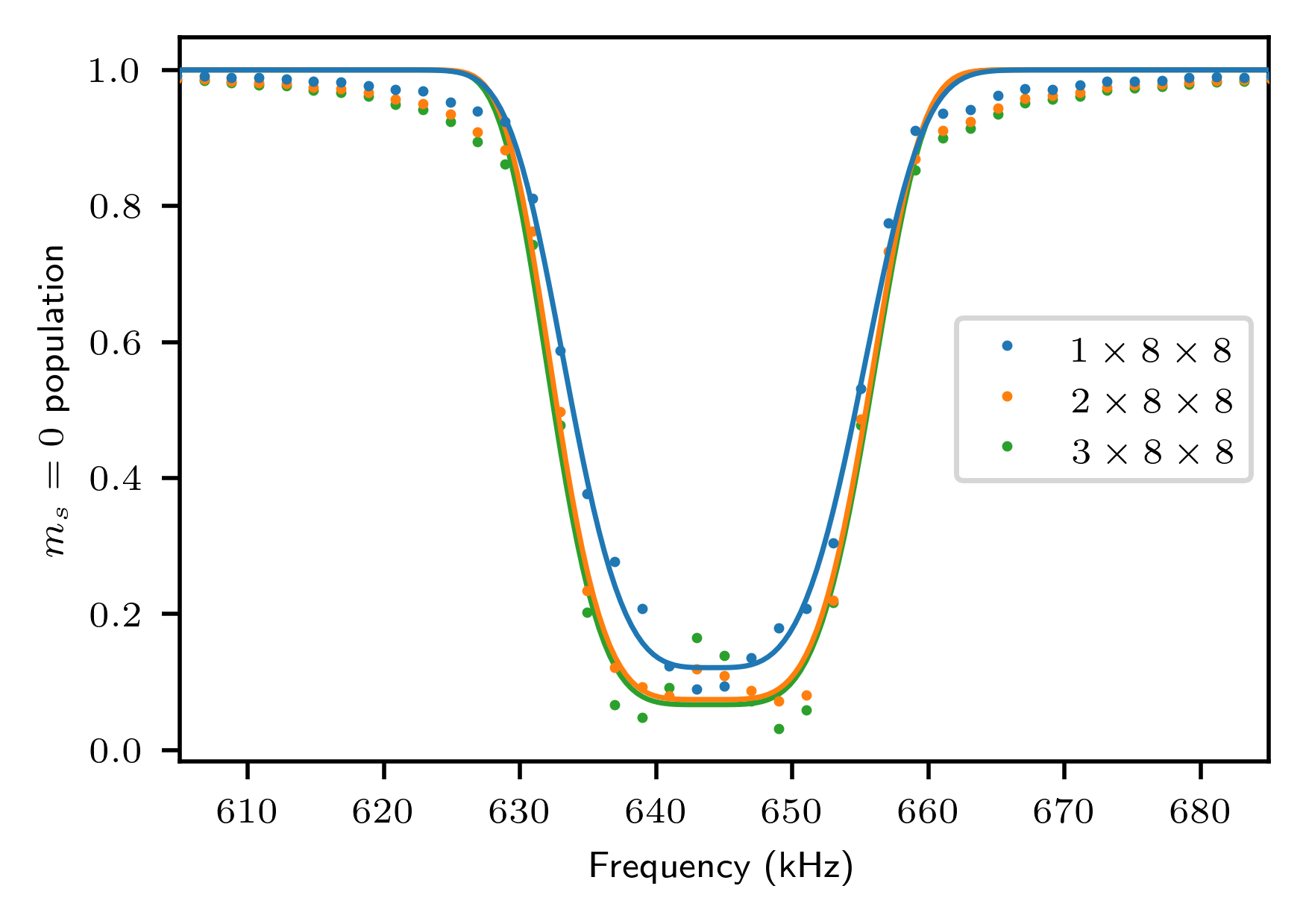}
    \caption{Comparison of the simulated AXY spectra for multiple spin layers.
    The simulated AXY spectrum of a \nv center at $1\,\text{nm}$ distance from a one-dimensional $8\times8$ nuclear spin layer (blue) is compared with the spectra of two (orange) and three (green) layers.
    The solid lines show the corresponding fit with a generalized normal distribution.
    }
    \label{fig:supplementary_fig_9}
\end{figure}

In addition to the tests described above, we investigate how the simulation results change when multiple spin layers are stacked.
\autoref{fig:supplementary_fig_9} shows the spectra for a $8\times8$ grid of \nuc nuclear spins for one (blue), two (orange) and three (green) stacked layers, each separated by the next-neighbor distance of $0.154\,\text{nm}$. 
The layers are placed at a distance of $1\,\text{nm}$ from the \nv center and solid lines show the corresponding fit with a generalized normal distribution~\cite{nadarajah2005generalized}. 
Compared to one layer, the variance for two layers changes by $3.78\,\%\pm0.12\,\%$ and by $6.82\,\%\pm0.13\,\%$ for three layers.
Although this might seem large at first, the difference in the distance estimation is within the margin of error.
Using the transfer function reported in the Methods section of the main text, we obtain a distance of $0.988\,\text{nm}\pm0.023\,\text{nm}$ for two layers and $0.981\,\text{nm}\pm0.023\,\text{nm}$ for three layers.
The impact of the layer thickness becomes smaller as the distance to the \nv center decreases.
Moreover, multiple layers overestimate the number of next-neighbors for the nuclear spins.
For those reasons, and to reduce computation time, we base our estimation on one layer only.

\section{Distance Estimation Results}

\autoref{tab:supplementary_sample_A_results} shows the results of the distance estimation for sample A and \autoref{tab:supplementary_sample_B_results} for sample B.
The reported variance errors correspond to the fit's standard deviation and the errors of the estimated distance are obtained through Gaussian error propagation. 
No \nuc resonance is observed for \nv centers No.\,1 and 3 in sample A, while in sample B, the spectrum of \nv center No.\,6 is too polluted to identify a clear resonance due to strong hyperfine interaction with nearby nuclear spins.
All other spectra exhibit a clear resonance that can be used to estimate the distance between the \nv center and the \nuc layer, provided that the variance extracted from the generalized normal distribution fit exceeds the estimation limit of 22.82\,ns${}^2$ (corresponding to a distance closer than 1.5\,nm).
\nv centers with variances below this threshold are marked by a "-" sign in the tables.
Within the measurable range, we obtain a mean distance of $1.01\,\text{nm} \pm 0.06\,\text{nm}$ for sample A and $1.066\,\text{nm} \pm 0.017\,\text{nm}$ for sample B, where the errors denote the mean's uncertainty.

\begin{table}[h]
    \renewcommand{\arraystretch}{1.25}
    \centering
    \caption{Fit results for sample A. The columns show the obtained variance $v$, its error $\Delta v$, the estimated distance $d$ and the corresponding error $\Delta d$. The errors reflect the fit's standard deviation, which is propagated using Gaussian error propagation.}
    \begin{tabular}{|c|S|S|S|S|}
         \hline
         \nv center & {$v$ (ns${}^2$)} & {$\pm \Delta v$ (ns${}^2$)} & {$d$ (nm)} & {$\pm\Delta d$ (nm)}\\
         \hline
         1 & 29.3       & 2.1   &  1.33     & 0.05      \\
         2 & 13.0       & 2.1   &  {-}      & {-}       \\
         3 &  {-}       & {-}   &  {-}      & {-}       \\
         4 &  20        & 2     &  {-}      & {-}       \\
         5 &  23        & 15    &  1.5      & 0.6       \\
         6 &  {-}       & {-}   &  {-}      & {-}       \\
         7 &  227       & 29    &  0.77     & 0.03      \\
         8 &  12        & 2     &  {-}      & {-}       \\
         9 &  631       & 4     &  0.631    & 0.022     \\
         10 &  46.4     & 1.8   &  1.130    & 0.027     \\
         11 &  0.7      & 0.5   &  {-}      & {-}       \\
         12 &  12.8     & 2.8   &  {-}      & {-}       \\
         13 &  9.0      & 1.9   &  {-}      & {-}       \\
         14 &  21       & 4     &  {-}      & {-}       \\
         15 &  5.9      & 2.1   &  {-}      & {-}       \\
         16 &  38       & 4     &  1.20     & 0.05      \\
         17 &  30.8     & 2.6   &  1.30     & 0.05      \\
         18 &  682.5    & 5     &  0.622    & 0.022     \\
         19 &  242.6    & 5     &  0.764    & 0.023     \\
         20 &  124.7    & 4     &  0.881    & 0.024     \\
         \hline
    \end{tabular}
    \label{tab:supplementary_sample_A_results}
\end{table}

\begin{table}[h]
    \renewcommand{\arraystretch}{1.25}
    \centering
    \caption{Fit results for sample B. The columns show the obtained variance $v$, its error $\Delta v$, the estimated distance $d$ and the corresponding error $\Delta d$. The errors reflect the fit's standard deviation, which is propagated using Gaussian error propagation.}
    \begin{tabular}{|c|S|S|S|S|}
         \hline
         \nv center & {$v$ (ns${}^2$)} & {$\pm \Delta v$ (ns${}^2$)} & {$d$ (nm)} & {$\pm\Delta d$ (nm)}\\
         \hline
         1 & 87.8   & 1.6   & 0.955     & 0.024 \\
         2 & 20.8   & 2.4   & {-}       & {-}   \\
         3 & 195    & 5     & 0.799     & 0.023 \\
         4 & 29.0   & 2.5   & 1.33      & 0.06  \\
         5 & 1287   & 14    & 0.552     & 0.021 \\
         6 & {-}    & {-}   & {-}       & {-}   \\
         7 & 146    & 8     & 0.851     & 0.025 \\
         8 & 62.4   & 1.6   & 1.040	   & 0.024 \\
         9 & 45.8   & 1.1   & 1.135     & 0.025 \\
         10 & 27.0  & 2.4   & 1.37      & 0.06  \\
         11 & 27.8  & 1.5   & 1.36      & 0.04  \\
         12 & 123   & 6	   & 0.884     & 0.025 \\
         13 & 23    & 4     & 1.50      & 0.14  \\
         14 & 23.9  & 2.5   & 1.46      & 0.09  \\
         15 & 21.1  & 2.8   & {-}       & {-}   \\
         16 & 79    & 4     & 0.981     & 0.025 \\
         17 & 26    & 8     & 1.4       & 0.2   \\
         18 & 82.8  & 2.4   & 0.968     & 0.024 \\
         19 & 76.4  & 2.5   & 0.988     & 0.025 \\
         20 & 1269  & 18	   & 0.553     & 0.021 \\
         \hline
    \end{tabular}
    \label{tab:supplementary_sample_B_results}
\end{table}


\section{Simulation of polarization dynamics}
\begin{figure}
    \centering
    \includegraphics[width=0.48\textwidth]{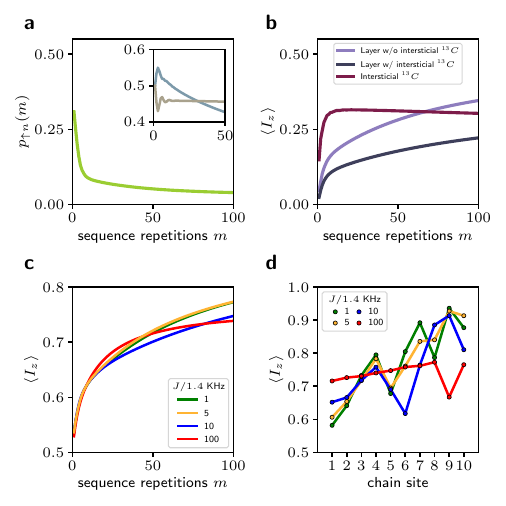}
    \caption{Polarization dynamics for the ten-nuclear spin array, with spin-locking time $\tau=5.0\,\text{\textmu s}$ and experimental waiting time $t_w=3.0\,\text{\textmu s}$. 
    \textbf{a)} Typical behavior of the \nv center for the NOVEL protocol, showing monotonous decay. We see a backflow of polarization during the protocol for particular values $\tau=1.9\,\text{\textmu s}$ and  $t_w=2.2\,\text{\textmu s}$ (blue line) and after the addition of a strongly coupled \nuc with coupling constants $A_{zx}\approx (2\pi)\,0.5\,\text{MHz}$ and $A_{zz}\approx (2\pi)\,0.3\,\text{MHz}$ (brown line). 
    \textbf{b)} Polarization blockade induced by the presence of the strongly coupled \nuc, showing the average polarization over the layer with and without such nuclei. 
    \textbf{c)} Average polarization of the nuclear spin chain as a function of the coupling strength between the constituent \nuc.
    \textbf{d)} Polarization distribution in the chain after $100$ repetitions of the sequence.}
    \label{fig:PolDynamics_supp1}
\end{figure}

\begin{figure}
    \centering
    \includegraphics[width=0.35\textwidth]{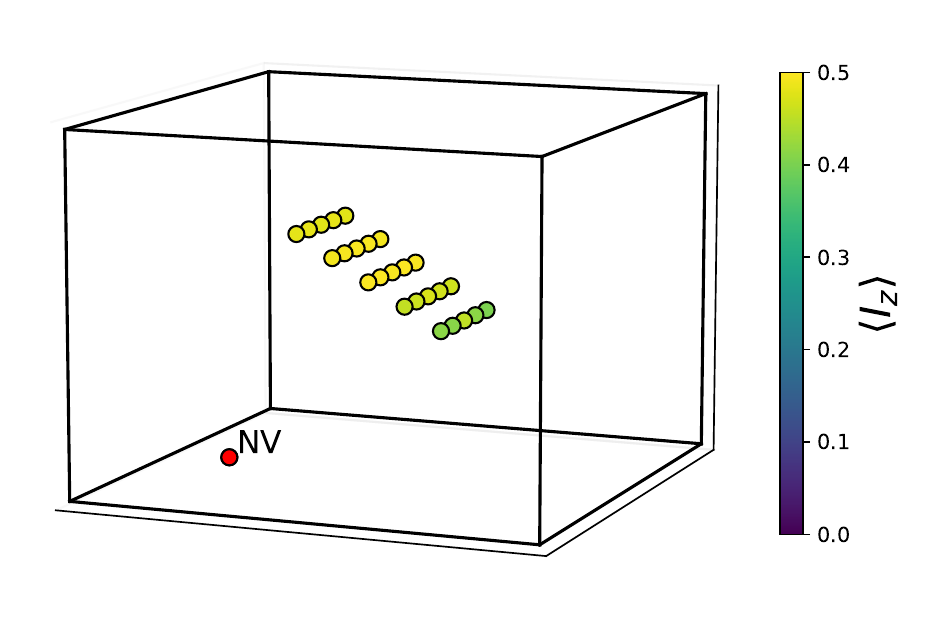}
    \caption{Polarization distribution of the two-dimensional layer for a spin-locking time $\tau=2.5\, \mu\text{s}$ and 2000 repetitions of the NOVEL protocol, where all \nuc nuclei show a polarization higher than $80\%$.}
    \label{fig:2DPolDistribution_supp2}
\end{figure}

The NOVEL polarization protocol consists of the system's unitary evolution under the effect of a mw drive satisfying the Hartmann-Hahn condition and of duration $\tau$ in between of two $\big(\frac{\pi}{2}\big)$ rotations (see Figure~2 in the main text) \cite{henstra1988nuclear}. 
The application of several repetitions of the sequence ensures an effective polarization transfer between the electron spin and a cluster of nuclear spins.
The NOVEL Hamiltonian, in the presence of inter-nuclear spin interaction is defined as
\begin{equation}
    H =   \Omega S_y+\gamma_N B_z\sum_{i} I_z^{i} + S_z \sum_i \Vec{A}_i \Vec{I} ^ i + H_{n-n} ,
    \label{hamiltonian}
\end{equation}
with $S_z=\frac{1}{2}(\sigma_z-\mathds{1}$), $S_y=\frac{1}{2}\sigma_y$\ spin operators acting on the electron spin and $I_i^j=\frac{1}{2}\sigma_i^j$ operators on the j-th nuclear spin. 
Under the rotating wave approximation (assuming $\gamma_N B \gg (\mu_0 \gamma_N^2)/(4\pi r_{ij}^3)$), the nearest-neighbor nuclear spin interaction Hamiltonian is given by~\cite{cai2013large}
\begin{equation}
    H_{n-n}=\sum_{<i,j>} J_{ij} \left( I_z^i I_z ^j - \frac{1}{2}(I_x^i I_x^j + I_y^i I_y^j)\right),
\end{equation}
with $J_{ij}=\frac{\mu_0}{4\pi}\frac{\gamma_N^2}{\hat{r}_{ij}^3}(1-3(r_{ij}^z)^2)$. 
To match the experimental conditions, we use a magnetic field of $B=600\,\text{G}$. 
We assess the polarization dynamics of the system by simulating a one-dimensional chain composed of ten interacting nuclear spins at $1$ nm distance from the \nv center, with equal internuclear coupling $J \approx (2\pi)\,1.4\,\text{kHz}$, assuming that all of them are separated $r\approx 1.54\,\text{\AA}$, the nearest neighbor distance in the diamond lattice.
Typically, the population of the \nv center $\ket{0}$ electron spin will decrease monotonically  as a signature of polarization transfer between the electron spin and the nuclear spins, shown in \autoref{fig:PolDynamics_supp1}\,a.
In agreement with experimental observations, we find that polarization backflow occurs under certain conditions during the initial stages of the polarization process. 
As numerically shown in \autoref{fig:PolDynamics_supp1}\,a, this backflow can be attributed to the presence of a strongly coupled nuclear spin and to free nuclear spin evolution during the experimental waiting time. 
We further suggest that the presence of a strongly coupled spin hinders the polarization transfer to the remaining \nuc atoms in the layer (\autoref{fig:PolDynamics_supp1}\,b), in line with recent reports \cite{whaites2023hyperpolarization,sasaki2024suppression}.\\
The existence of internuclear interaction suggests the need to study the polarization redistribution. 
We analyze its effect by numerical simulations on the same ten-nuclear spin array, showing in \autoref{fig:PolDynamics_supp1}\,c and \autoref{fig:PolDynamics_supp1}\,d  how increasing the interaction strength reduces the amount of polarization that is transferred to it for that timescale. 
The interaction also diffuses the polarization, tending to homogenize the distribution~\cite{karabanov2015dynamic}. Even so, for realistic nuclear spin coupling strengths $J \leq (2\pi)\,5$\,kHz, the effect can be neglected as the timescale of the inter-nuclear interaction is much longer than the pulse sequence time.\\
In \autoref{fig:2DPolDistribution_supp2}, we present the polarization distribution for the 25 closest nuclear spins of a two-dimensional layer at $1\,\text{nm}$ from the \nv center. 
It shows saturation at more than $80\%$ for all the nuclear spins after 2000 repetitions of the protocol for a spin-locking time $\tau=2.5\,\mu \text{s}$.
%

\section{Readout via NOVEL}

\begin{figure}
    \centering
    \includegraphics[width=0.48\textwidth]{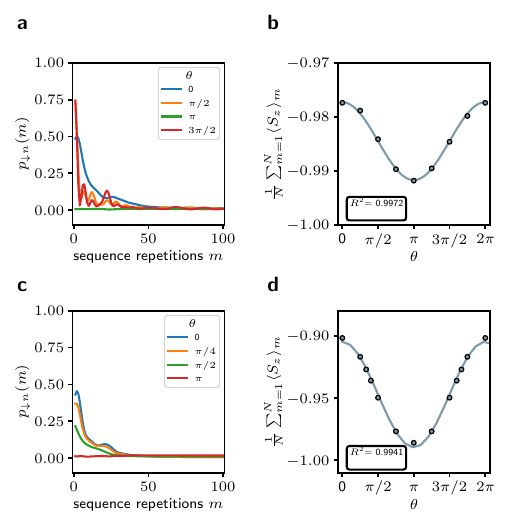}
    \caption{Readout of layer polarization via NOVEL, showing full polarization dynamics and the fitting to the expected cosine in both studied configurations. 
    \textbf{a)}, \textbf{b)} Nuclei initially polarized and then rotated an angle $\theta$.
    \textbf{c)},\textbf{d)} Spins in thermal state with initial polarization controlled through the parameter $\theta$.}
    \label{fig:Readout_NOVEL_supp3}
\end{figure}

Any arbitrary experiments starts by polarizing the nuclear spins into $\vert{\uparrow_n}\rangle$ via the NOVEL sequence.
Afterwards, the actual experiment on the nuclear spin layer is performed.
Changes of the nuclear spin states due to, e.g., spin diffusion or rotations induced by rf fields, will result in different curves in each case.
Polarizing the layer to the opposite state, i.e. reverting the initially built up polarization with the NOVEL sequence, allows to read out the remaining polarization in the nuclear spin environment~\cite{scheuer2017robust,unden2018coherent}.
We infer the collective nuclear $\langle I_z\rangle$ expectation value from the measurements by first calculating the ratio $R_n$ from the summed signals of both curves according to
\begin{equation}
    R_n= \frac{\sum_{m=1}^{M}p_{\downarrow_n}\left(m\right)}{\sum_{m=1}^{M}p_{\uparrow_n}\left(m\right)}.
    \label{eq:readout}
\end{equation}
The amplitude $A_{\text{Rabi}}$ and offset $R^0_{\text{Rabi}}$ from a nuclear Rabi fit can then be used to calculate the expectation value
\begin{equation}
    \langle I_z \rangle = \frac{R_n - \left(R^0_{\text{Rabi}} - A_{\text{Rabi}}\right)}{2A_{\text{Rabi}}} - \frac{1}{2}.
    \label{eq:readout_normalization}
\end{equation}
Changing the total number of applied NOVEL repetitions $M$ in the experiment or during the data analysis in post-processing allows to probe different parts of the dynamics within the layer.
For example, for a \nv center with an interstitial \nuc nuclei or an \nv center in close proximity to the \nuc layer, the first data points of the NOVEL measurement are dominated by the strongly coupled nuclear spins.
Thus, for small $M$, the computed collective $\langle I_z \rangle$ expectation value reflects mainly the dynamics of the nearest nuclear spins.
Conversely, for large $M$ the dynamics of distant nuclear spins become more pronounced. \\
To investigate the validity of our readout scheme described by \autoref{eq:readout}, we use the same cluster of \nuc nuclear spins at a distance of $1\,\text{nm}$ from the \nv center of the previous section and manually set all nuclear spin states to $\vert{\uparrow_n}\rangle$.
The spins are then rotated by the angle $\theta$ with $\theta\in[0,2\pi]$ around $I_y$ with the goal of reconstructing the initial nuclear spin state through the NOVEL sequence according to \autoref{eq:readout} and \autoref{eq:readout_normalization}.
Similar to the experiment, we choose the sequence depicted in Figure~2 in the main text for the readout, using $-\big(\frac{\pi}{2}\big)$ pulses.
We use a spinlock duration of 5\,\textmu s, a magnetic field of $B=600\,\text{G}$ and an amplitude matching $\gamma_nB$ with up to $M=500$ repetitions.
The obtained mean $\langle S_z\rangle$ values for different $\theta$ are then fitted by the expected $a\cdot\cos(\theta)+b$.
In \autoref{fig:Readout_NOVEL_supp3} a), we show the full evolution of $\langle S_z\rangle$ for different values of the rotation angle $\theta$ while in \autoref{fig:Readout_NOVEL_supp3} b), we show the fitting of the mean $\langle S_z\rangle$ with the expected cosine.\\
To further investigate the polarization readout, we now perform the same analysis with nuclear spins initially as a thermal state with a certain degree of polarization. To do so, we set the initial density matrix of each spin to be $\rho_n(0) = \frac{1}{2}(\mathds{1}+\cos(\theta)\sigma_z)$, controlling the initial polarization through the parameter $\theta$ as $\langle I_z(0)\rangle =\cos(\theta)$. We choose this form of the initial density matrix as its expectation value coincides with the first scenario.  In \autoref{fig:Readout_NOVEL_supp3} c) and d), we show a similar behavior as in the previous case, where the signal fits the expected cosine. Furthermore, the difference in signal between different polarization degrees is enlarged, in a better agreement with the experimental measurements.\\
Experimentally, the simulated signal corresponds to a Rabi measurement that is found to match well with the expected cosine.
We can thereby conclude that we can read out their collective polarization correctly via NOVEL.
Due to the large number of nuclear spins in our \nuc layer, the signal strength of the observed mean $\langle S_z \rangle$ is significantly larger than for the simulated, small cluster. \\
Simulations have been performed by mapping the system to a one-dimensional spin-$1/2$ chain with long range interactions. 
For this task, we used the ITensor Library~\cite{fishman2022itensor} in Julia. 
The system's initial state was defined as a Matrix Product Operator (MPO) as the nuclear spin bath is initially in a thermal state, here the fully mixed state as we work at room temperature ($kT\gg\hbar \omega$), and performed Time-Evolving Block Decimation (TEBD). 
The error in the simulations has been controlled by setting a cut-off in the smallest singular value of the MPO decomposition, letting the system achieve very large bond dimensions as the computational cost of this method is increased when we include long-range interactions. 
We reduce the cut-off until we see convergence in the measured magnetization.
Weak coupling between the \nv center and the nuclear spins makes the simulations tractable and reliable for the used timescales and system sizes.
DNP protocols present a decrease in entanglement after each reinitialization of the system, reaching a maximum value that starts to decrease after a particular number of repetitions, making the simulations when getting close to the non-equilibrium steady state.
%

\section{Spin properties of the \nuc layer}
\label{sec:supplementary_spin_properties}

\begin{figure*}
    \centering
    \includegraphics[width=0.98\textwidth]{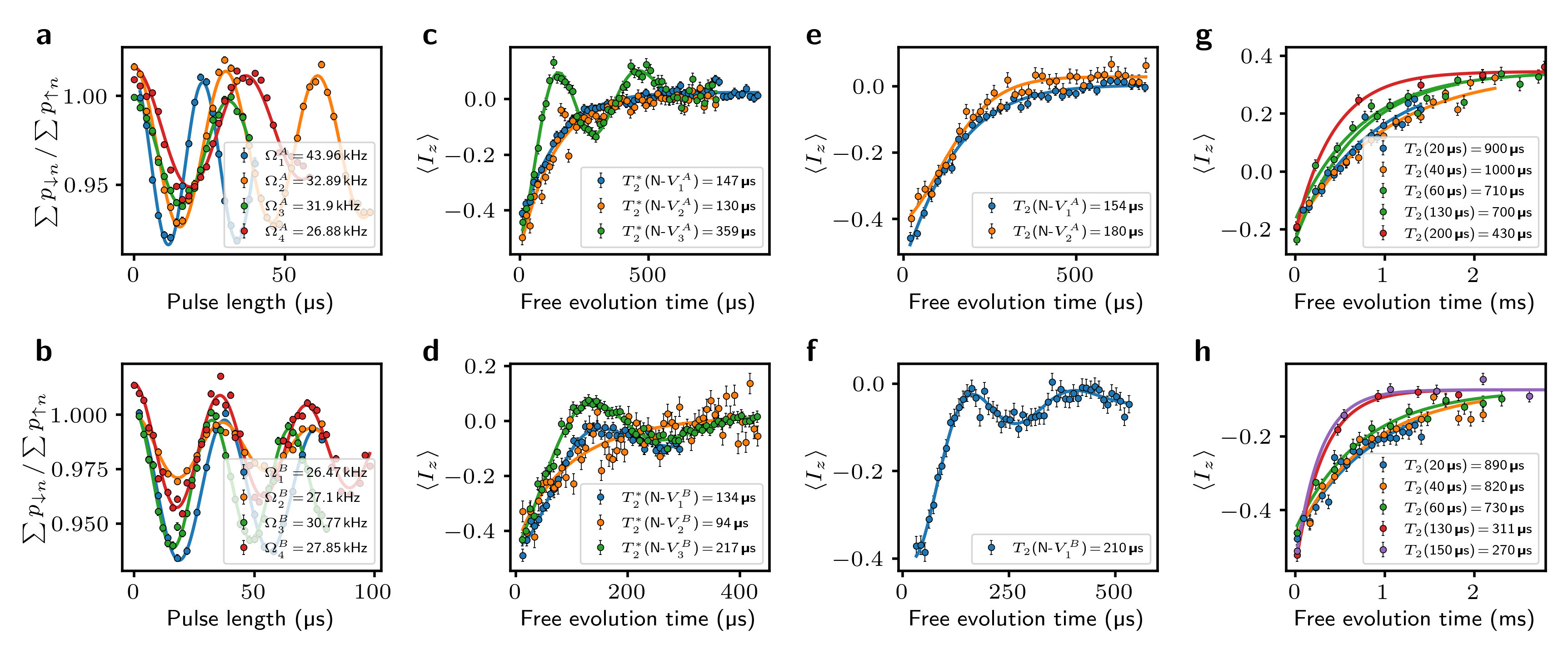}
    \caption{Coherent control of the \nuc layer. 
    Rabi measurements of different \nv centers for \textbf{a} sample A and \textbf{b} sample B.
    The solid lines show the corresponding fit result.
    \textbf{c} Ramsey measurements for sample A and \textbf{d} sample B.
    The colors assigned to the \nv centers are consistent across all sub-plots. 
    \textbf{e} Hahn Echo measurements for sample A and \textbf{f} sample B.
    \textbf{g} WAHUHA measurements of N-$V_2^A$ for sample A, read out with a $-\pi/2$ pulse and \textbf{h} WAHUHA measurements of N-$V_1^B$ for sample B, which is read out with a $\pi/2$ pulse instead.
    The different readout pulses in combination with imperfect state preparation lead to a different baseline~\cite{haeberlen2012high}.
    The colors correspond to different pulse spacings $\tau$, which are displayed by the argument of $T_2(\tau)$.}
    \label{fig:supplementary_fig_2}
\end{figure*}

\autoref{fig:supplementary_fig_2}\,a shows Rabi measurements of the \nuc layer for different \nv centers in sample A and \autoref{fig:supplementary_fig_2}\,b for sample B.
We obtain Rabi frequencies ranging from $(2\pi)\,(26.47\pm0.07)\,\text{kHz}$ to $(2\pi)\,(43.96\pm0.08)\,\text{kHz}$, depending on their distance to the copper wire, which is used to apply the microwave (mw) and rf control fields.
Although higher Rabi frequencies are achievable, the chosen parameters allow for multiple rf pulse gates without excessive heating of the wire used to apply the microwave and rf fields (see \autoref{sec:supplementary_setup}).
Increased Rabi frequencies introduce more heat into the system, potentially causing loss of confocal focus and breaking of the wire.
Nevertheless, for sequences requiring a large number of pulses, e.g., WAHUHA, heating remains a limiting factor.
Microwave structures fabricated directly on the diamond surface or external microwave coils may help alleviate these limitations. \\
The pulse sequences of all presented experiments are shown in the main text. \\
We measure the nuclear dephasing time $T_2^*$ in a Ramsey experiment.
The dephasing times are extracted by fitting the signal's envelope with a stretched exponential decay.
For sample A we obtain
\begin{itemize}
    \item $T_2^*(\text{N-}V_1^A)=147\,\text{\textmu s} \pm 7\,\text{\textmu s}$,
    \item $T_2^*(\text{N-}V_2^A)=130\,\text{\textmu s} \pm 10\,\text{\textmu s}$,
    \item $T_2^*(\text{N-}V_3^A)=359\,\text{\textmu s} \pm 19\,\text{\textmu s}$,
\end{itemize}
and for sample B:
\begin{itemize}
    \item $T_2^*(\text{N-}V_1^B)=134\,\text{\textmu s} \pm 9\,\text{\textmu s}$,
    \item $T_2^*(\text{N-}V_2^B)=94\,\text{\textmu s} \pm 18\,\text{\textmu s}$,
    \item $T_2^*(\text{N-}V_3^B)=217\,\text{\textmu s} \pm 12\,\text{\textmu s}$.
\end{itemize}
While the size of the data set does not permit a quantitative comparison, both samples exhibit dephasing times in the low hundreds of microseconds. 
This is significantly shorter than typical $T_2^*$ values of single \nuc nuclear spins at similar distances to the \nv center in natural abundance diamond, which are mainly limited by the \nv center's $T_1$~\cite{pfender2017nonvolatile}.
The short dephasing times thus indicate a strong interaction among the nuclear spins.
Some of the Ramsey measurements show oscillations which can be either attributed to these interactions or to detuning.
In our NMR reference measurements (see \autoref{sec:supplementary_nmr_measurements}) with 100\,\% \nuc diamond powder, we observe an even shorter $T_2^*$ with a small oscillatory component. 
Apart from inhomogeneities in the \nuc concentration, the difference can be attributed to our \nuc layer being only a few atomic layers thick, severely limiting the possible nuclear interactions in comparison to the \nuc powder. 
In contrast to \nuc crystal, all orientations contribute to the signal of the powder, damping the oscillations and reducing the dephasing time. \\
As we normalize the data by the corresponding Rabi amplitude to calculate the collective $\langle I_z\rangle$ expectation value, temperature fluctuations can negatively impact the normalization.
Such temperature shifts can shift the \nv centers resonance frequency as well as the static magnetic field, detuning the Hartmann-Hahn resonance.
The obtained NOVEL polarization curves are then slightly altered, leading to errors in the normalization.
This can, e.g., lead to an offset in the observed decay with a decay level $\langle I_z\rangle \ne 0$. 
A more detailed discussion of normalization errors can be found in \autoref{sec:normalization_errors}.\\
Assuming the dipolar interactions among the nuclear spins are the main source of decoherence, spin echo sequences, like the Hahn echo, should not be able to prolong the coherence time or refocus the observed oscillations~\cite{choi2020robust}. 
The corresponding Hahn echo measurements are shown in \autoref{fig:supplementary_fig_2}\,e for sample A and in \autoref{fig:supplementary_fig_2}\,f for sample B.
We again fit the signal's envelope with an stretched exponential decay, leading to 
\begin{itemize}
    \item $T_2(\text{N-}V_1^A)=154\,\text{\textmu s} \pm 9\,\text{\textmu s}$
    \item $T_2(\text{N-}V_2^A)=180\,\text{\textmu s} \pm 10\,\text{\textmu s}$
\end{itemize}
for sample A and
\begin{itemize}
    \item $T_2(\text{N-}V_1^B)=210\,\text{\textmu s} \pm 40\,\text{\textmu s}$
\end{itemize}
for sample B, which in all cases only show minor improvements over their respective dephasing times.
\autoref{fig:supplementary_fig_2}\,f also shows non-refocused oscillations, likely arising from the dipolar interactions among the nuclear spins. \\
Homonuclear decoupling sequences like WAHUHA~\cite{waugh1968approach, balasubramanian2019dc, choi2020robust} on the other hand, are designed to decouple such dipolar interactions. 
The respective measurements are shown in \autoref{fig:supplementary_fig_2}\,g for N-$V_2^A$ in sample A and in \autoref{fig:supplementary_fig_2}\,h for N-$V_1^B$ in sample B.
The different colors represent different pulse spacings $\tau$. 
We again observe similar dynamics for both samples, which exhibit a significant increase of their coherence times for short $\tau$, confirming strong dipolar interactions among the nuclear spins as the main source of decoherence. \\
Short $\tau$ require more sequence repetitions and thus more pulses to measure the coherence at long free evolution times.
Due to heating, we are unable to capture the entire decay for such short $\tau$.
During the fitting procedure we therefore assume a common baseline, using the decay level of the fit with largest $\tau$.  
The data is fitted with an exponential decay.
Since the WAHUHA sequence scales down all $I_z$ interactions by a factor of $1/\sqrt{3}$~\cite{haeberlen2012high}, the x-axes in \autoref{fig:supplementary_fig_2}\,g, \autoref{fig:supplementary_fig_2}\,h and our fit results are adjusted accordingly. \\
Additionally, since the nuclear spins are not initialized along the effective direction of the WAHUHA's toggling frame, a persistent longitudinal magnetization remains~\cite{haeberlen2012high}.
In Sample A (\autoref{fig:supplementary_fig_2}\,g), the population is read out with a $-\pi/2$ pulse, resulting in a different baseline~\cite{haeberlen2012high}. \\

\begin{figure}
    \centering
    \includegraphics[width=0.48\textwidth]{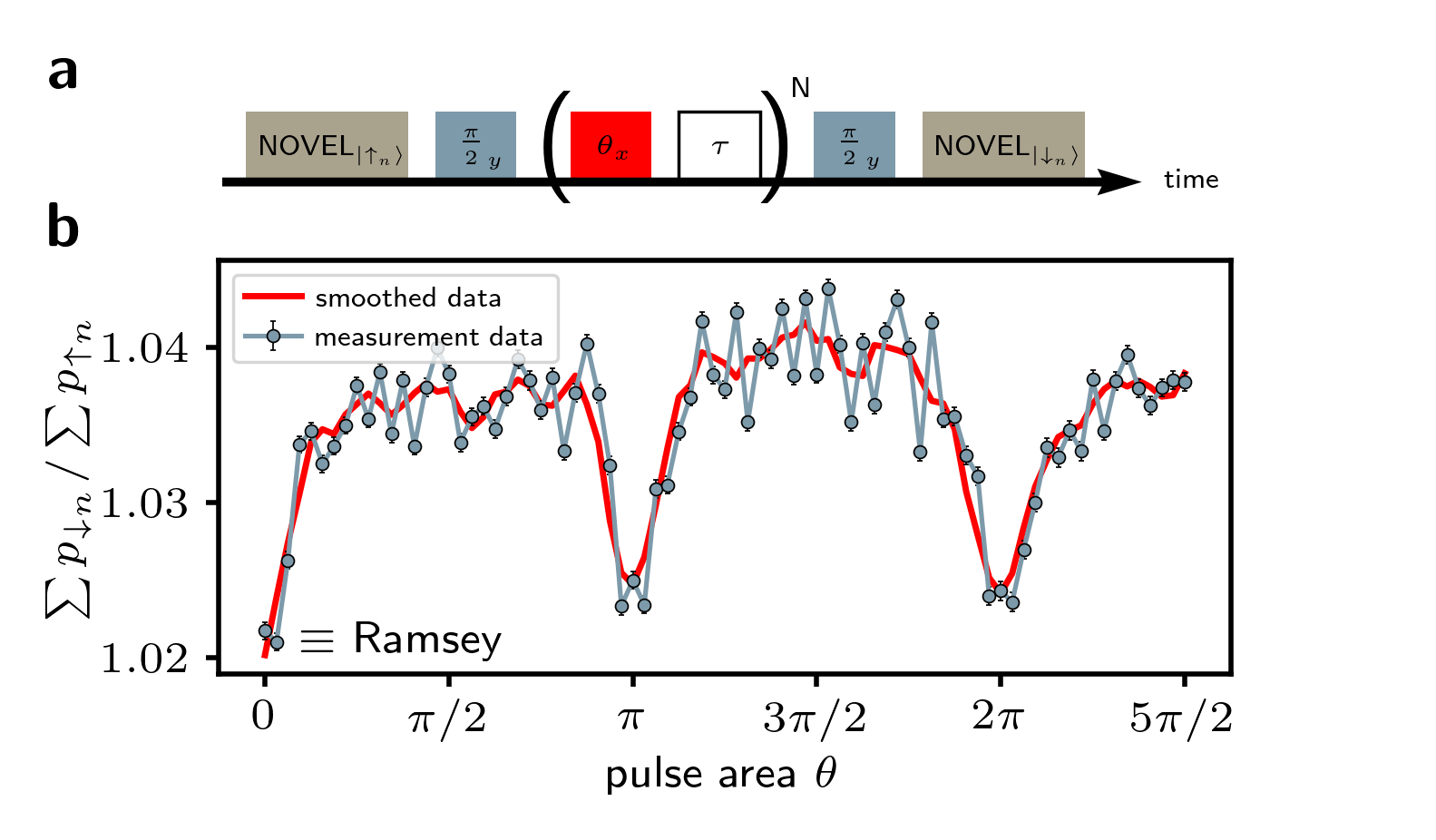}
    \caption{\nuc decoupling for different pulse areas.
    \textbf{a} Pulse sequence to probe the dipolar interaction in the nuclear spin layer.
    \textbf{b} Measurement result for different pulse areas $\theta$ with $N=7$ and $\tau=20\,\text{\textmu s}$.
    The red line shows the smoothed data points (blue).
    }
    \label{fig:supplementary_fig_3}
\end{figure}
Another method of testing the dipolar interaction between the nuclear spins is by employing the pulse sequence sketched in Fig.~\ref{fig:supplementary_fig_3}\,a from~\cite{ajoy2020dynamical, beatrez2021floquet}.
While it is not feasible to measure the full decay for every pulse area $\theta$ in our system due to the long measurement time, we measure the coherence for fixed $N=7$ and a pulse spacing of $\tau=20\,\text{\textmu s}$.
For $\theta=0$ the obtained signal corresponds to a Ramsey measurement.
An increase of the signal directly translates to an increase of the coherence lifetime.
To compensate for the poor signal-to-noise ratio, we smooth the data as guidance for the eye (Savitzky-Golay filter with a window length of eight data points and a polynomial order of three).
For $\theta=\pi$, we observe a steep decrease of the coherence, which is expected as the dipolar interactions among the spins are the leading source of decoherence and are not suppressed for this pulse area.
%

\section{NMR Measurements}
\label{sec:supplementary_nmr_measurements}

\begin{figure}[h]
    \centering
    \includegraphics[width=0.48\textwidth]{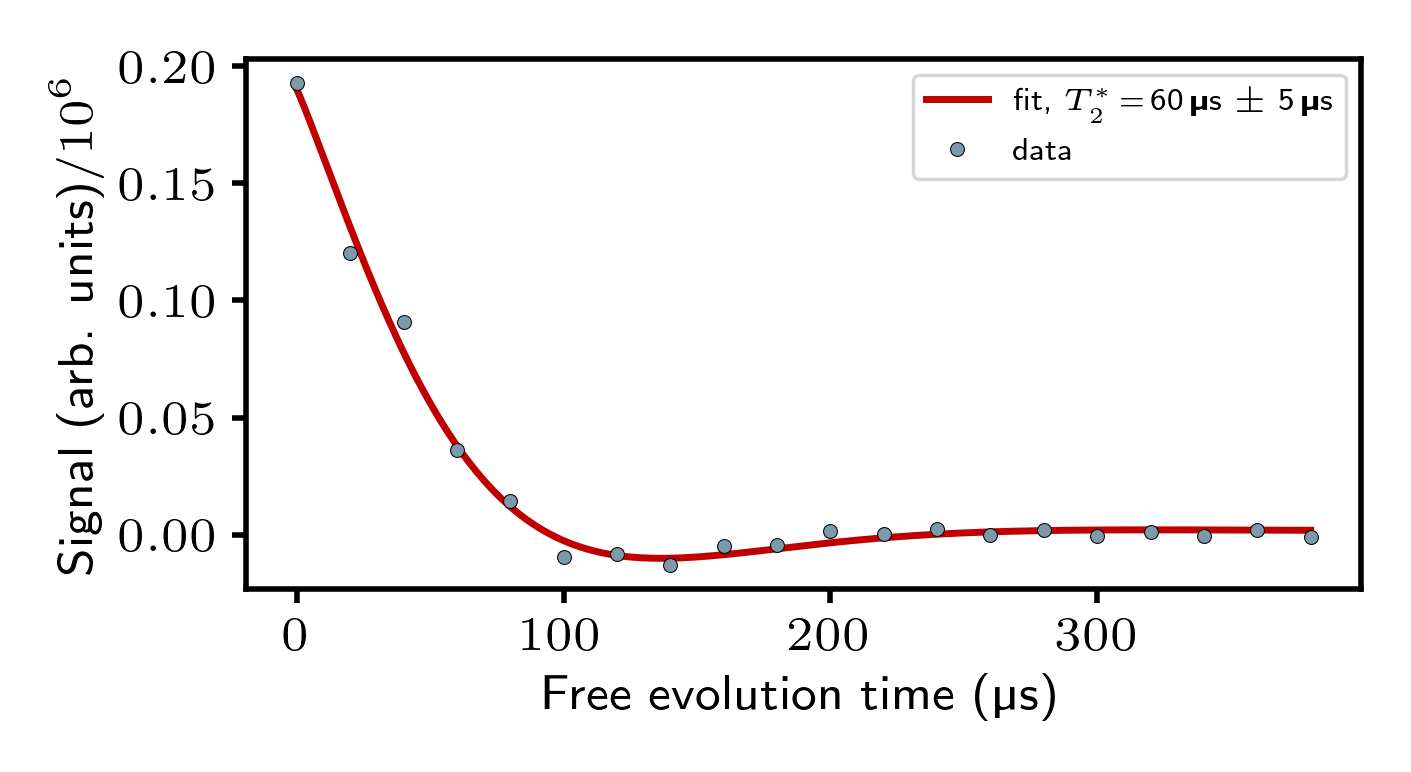}
    \caption{Free induction decay for $\sim 100\%$ $^{13}$C-enriched diamond powder and fit (the data was measured  at \SI{7.05}{\tesla} magnetic field with conventional NMR).  }
    \label{fig:supplementary_nmr}
\end{figure}

\autoref{fig:supplementary_nmr} shows the free induction decay measured at \SI{7.05}{\tesla} magnetic field for diamond powder with nearly $100\%$ $^{13}$C content ($\sim 96\%$ $^{13}$C, as in~\cite{mindarava2021coreshell}). 
The fit yields a dephasing time of $T_2^* = \SI{60\pm5}{\micro\second}$.
The experiments are performed with a $300\,\text{MHz}$ WB NMR with an Avance III console (Bruker BioSpin).

\section{Discrete Time-Crystalline order}

\subsection{Coherence Properties of the \nuc Nuclear Spins}

\begin{figure}
    \centering
    \includegraphics[width=0.48\textwidth]{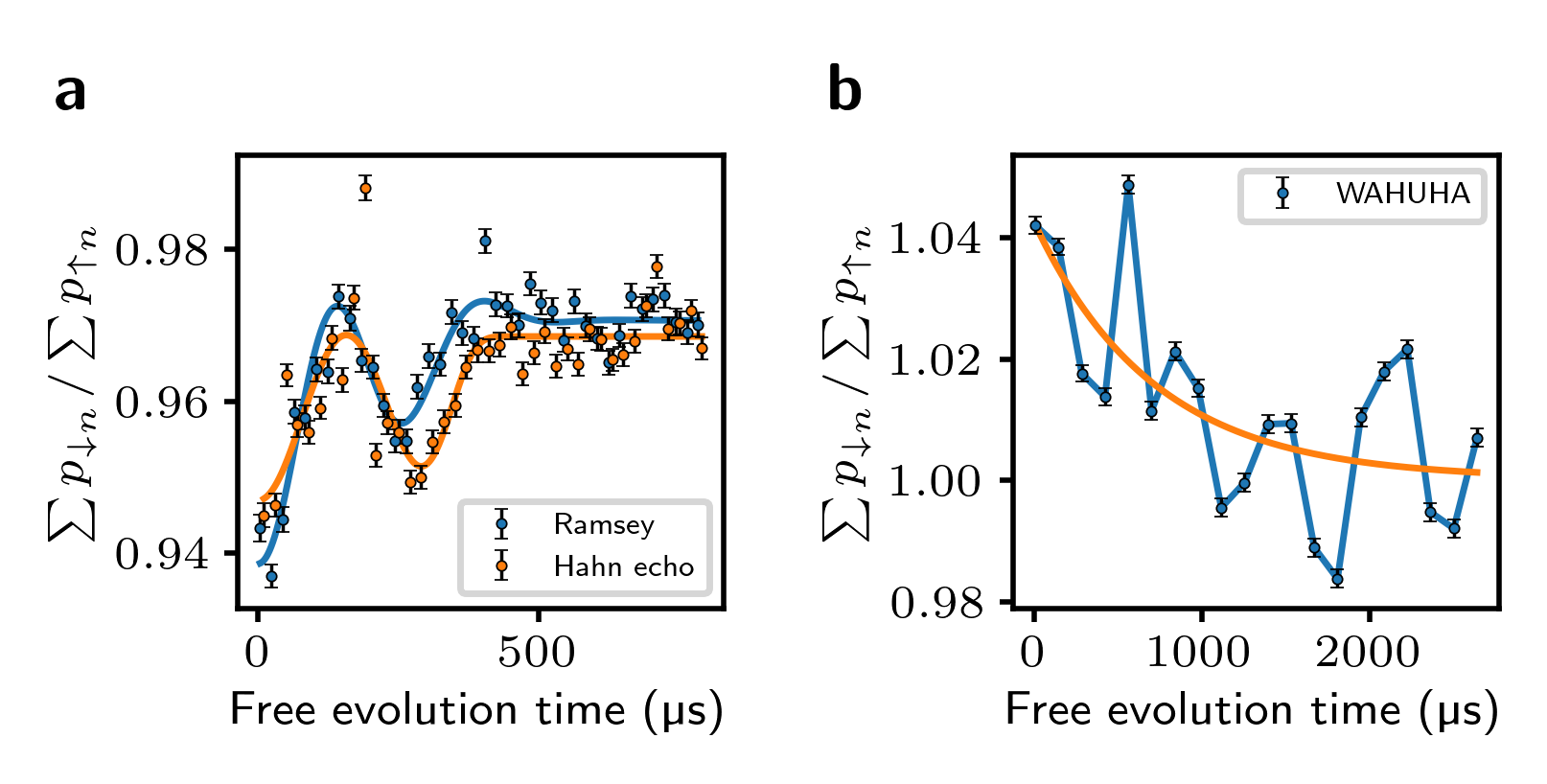}
    \caption{Coherence properties of the \nuc nuclear spins.
    \textbf{a} Ramsey measurement (blue) and Hahn echo measurement (orange) on the nuclear spins.
    The solid line shows the corresponding fit result, yielding $T_2^*=324\,\text{\textmu s} \pm 16\,\text{\textmu s}$ and $T_2=360\,\text{\textmu s} \pm 14\,\text{\textmu s}$.
    \textbf{b} WAHUHA measurement. 
    The fit with an exponential decay (orange line) yields a coherence time of $T_2=720\,\text{\textmu s} \pm 60\,\text{\textmu s}$.
    }
    \label{fig:supplementary_fig_12}
\end{figure}

We start by investigating the coherence properties of the employed \nuc nuclear spins.
Therefore we measure the nuclear dephasing time in a Ramsey experiment, shown in \autoref{fig:supplementary_fig_12}\,a, yielding a $T_2^*$ of $324\,\text{\textmu s} \pm 16\,\text{\textmu s}$.
The Hahn echo, shown as the orange line in \autoref{fig:supplementary_fig_12}\,a, yields almost the same $T_2$ of $360\,\text{\textmu s} \pm 14\,\text{\textmu s}$, which is consistent with the observations presented in \autoref{sec:supplementary_spin_properties}.
Since the Hahn echo shows no significant improvement, the coherence time is likely limited by spin-spin interactions among the nuclear spins.
\autoref{fig:supplementary_fig_12}\,b shows the WAHUHA experiment for a pulse spacing of $\tau=40\,\text{\textmu s}$.
Although the measurement data appears noisy, the non-Markovian revivals/ oscillations clearly exceed the noise floor, indicating a coherence time much longer than the extracted $T_2$ of $720\,\text{\textmu s} \pm 60\,\text{\textmu s}$ from the exponential fit.
The signal may be heavily undersampled and requires further detailed characterization to identify its origin.
Nevertheless, the measurement confirms strong interactions among the nuclear spins.

\subsection{Influence of Markovian Dephasing}
\label{sec:influence_markovian_dephasing}

\begin{figure*}
    \centering
    \includegraphics[width=0.99\textwidth]{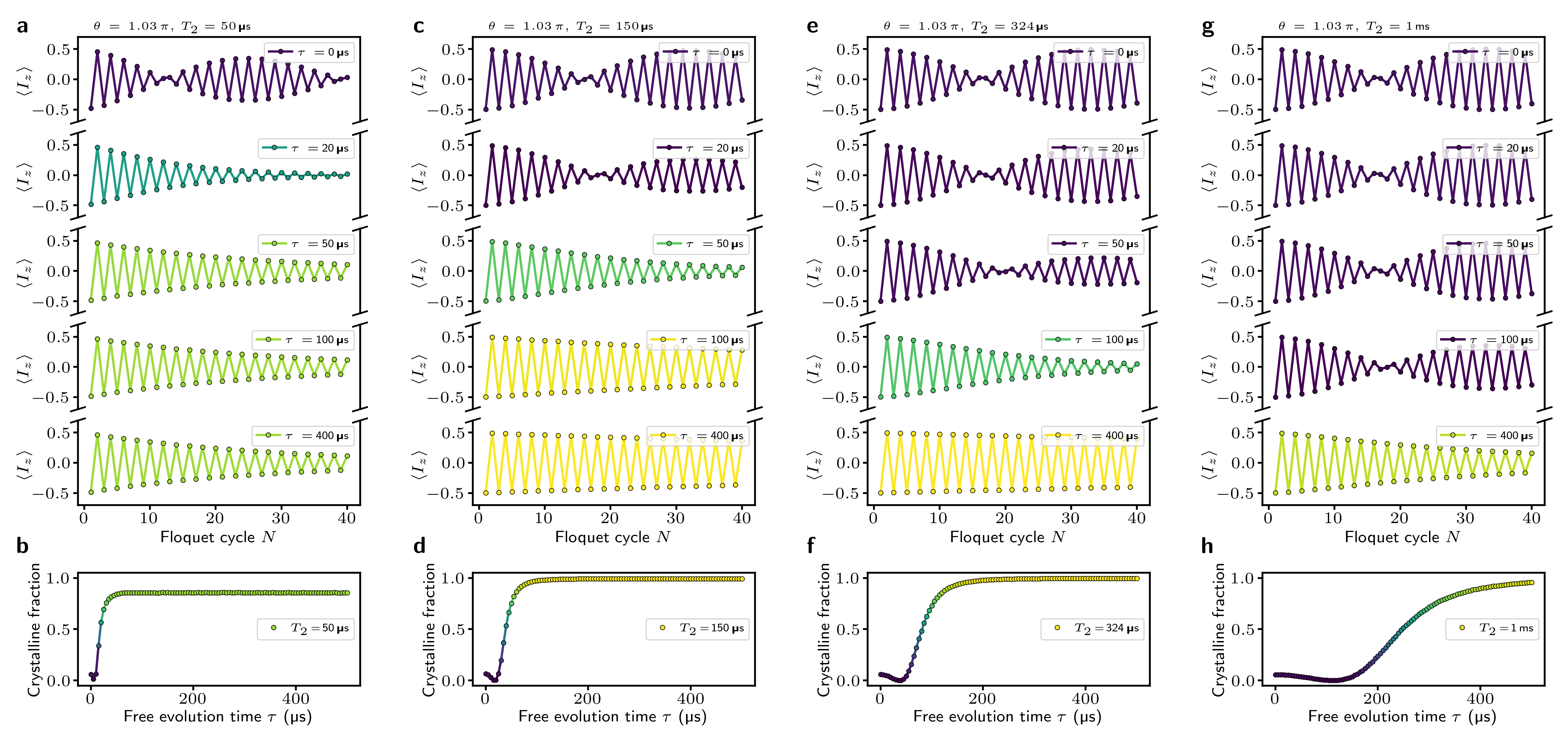}
    \caption{Simulated discrete time-crystal measurements for for a single \nuc nuclear spin with different Markovian dephasing times $T_2$.
    Each row shows the time trace for a different pulse spacing $\tau$ and a pulse area of $\theta=0.9\,\pi$.
    \textbf{a} Time traces for $T_2=50\,\text{\textmu s}$ and
    \textbf{b} achieved crystalline fraction as function of $\tau$.
    \textbf{c} \&
    \textbf{d} data for $T_2=150\,\text{\textmu s}$.
    \textbf{e} \&
    \textbf{f} data for $T_2=324\,\text{\textmu s}$ and
    \textbf{g} \&
    \textbf{h} data for $T_2=1\,\text{ms}$. 
    }
    \label{fig:supplementary_fig_6}
\end{figure*}

Markovian dephasing of the nuclear spins can lead to signals that may be mistakenly interpreted as signatures of discrete time-crystalline order~\cite{choi2017observation}.
To visualize this problem, we simulate the discrete time-crystal (DTC) measurements for a single \nuc nuclear spin for multiple Markovian dephasing times $T_2$.
The Markovian dephasing is modeled by a randomized detuning $\Delta = \sqrt{2}/T_2$~\cite{grimm2025coherent}.
The detuning is randomized independently for each free evolution time $\tau$ in the sequence, and the resulting state after $\tau$ is then obtained by averaging over 2000 repetitions of that free evolution time.
The pulse sequence itself is only applied once.
The detuning is randomized for each free evolution time individually.
Equivalent to the experiment, we choose rotation pulses with a 3\,\% over-rotation ($\theta=1.03\,\pi$) and a Rabi frequency of $(2\pi)\,37.14\,\text{kHz}$.
The results are shown in \autoref{fig:supplementary_fig_6}, where each column corresponds to a different $T_2$.
The maximum crystalline fraction (last row) increases with increasing dephasing time, and longer dephasing times also result in a slower, more gradual growth as $\tau$ increases. 
Thus, by only looking at the crystalline fraction, it is possible to mistakenly interpret these dynamics as the signature of a discrete time-crystalline order. \\
\autoref{fig:supplementary_fig_6}\,e shows the results for $T_2=324\,\text{\textmu s}$, matching the dephasing time of the nuclear spins.
Even when assuming such a short $T_2$, the simulation does not match the experimental observations.
From the WAHUHA measurements we know that the Markovian dephasing time is much longer.
\autoref{fig:supplementary_fig_6}\,g shows the simulation with an increased dephasing time of $1\,\text{ms}$.
With longer dephasing times, the agreement between simulation and measurement data further decreases.

\begin{figure}
    \centering
    \includegraphics[width=0.48\textwidth]{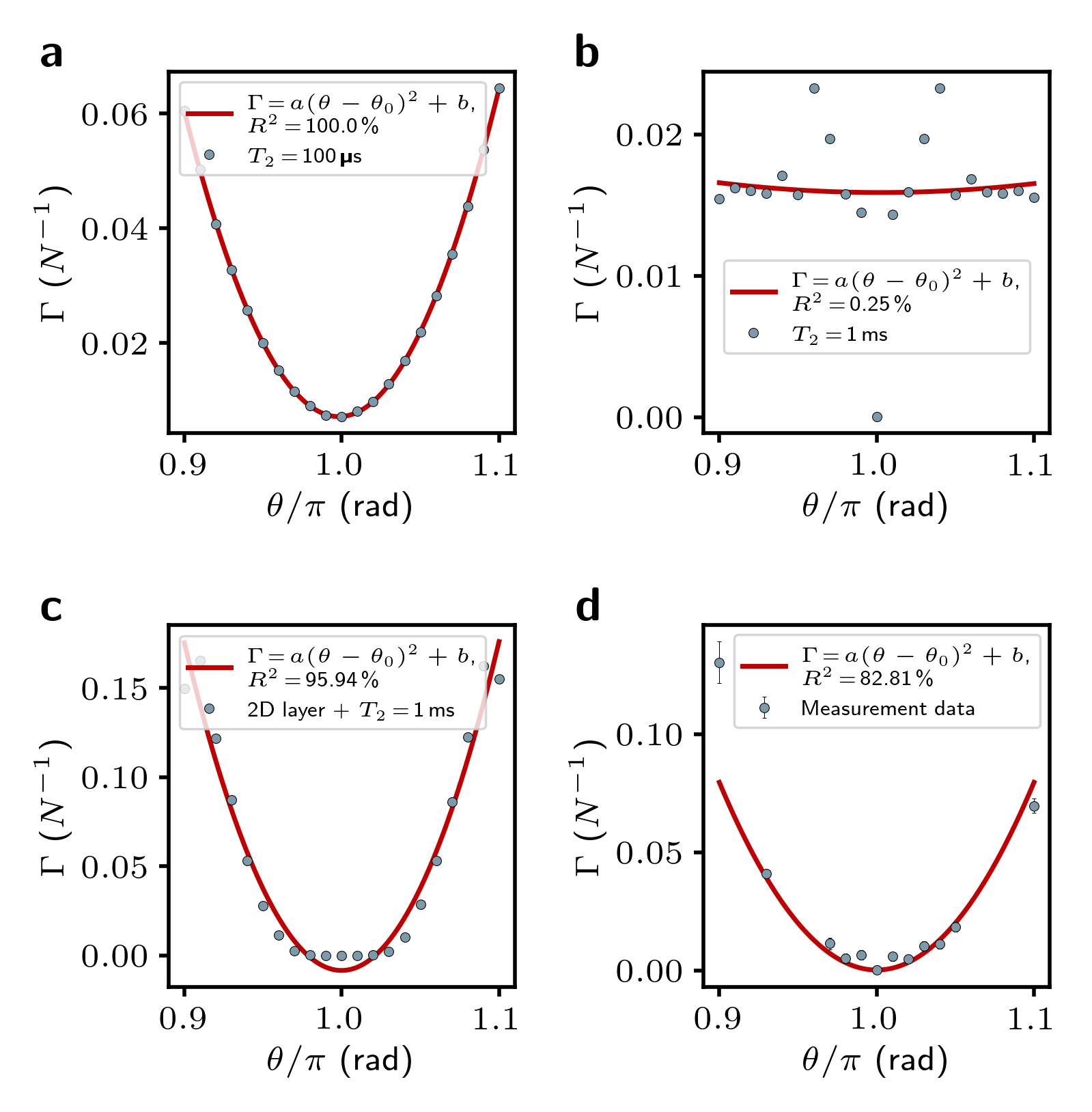}
    \caption{Dependency of the Markovian decay rate $\Gamma$ on the pulse area $\theta$.
    \textbf{a} Simulation results of a single \nuc spin for $T_2=100\,\text{\textmu s}$ and $\tau=125\,\text{\textmu s}$.
    The red line shows the quadratic fit.
    All decay rates are extracted by an stretched exponential fit of the first 40 Floquet cycles. 
    \textbf{b} For a much longer dephasing time of $T_2=1\,\text{ms}$, the quadratic dependency vanishes.
    \textbf{c} Simulation results of a 2D layer of nine interacting nuclear spins that are subject to Markovian dephasing with $T_2=1\,\text{ms}$.
    \textbf{c} Measurement data.
    }
    \label{fig:supplementary_fig_7}
\end{figure}

To further exclude Markovian dephasing as the source of the observed time-crystalline order, we measure the decay rate $\Gamma$ (in units of Floquet cycles) for different $\theta$.
Sufficiently strong Markovian dephasing leads to a quadratic dependency on $\theta$~\cite{choi2017observation}, as visualized in \autoref{fig:supplementary_fig_7}\,a for a pulse spacing of $\tau=125\,\text{\textmu s}$ and $T_2=100\,\text{\textmu s}$.
For long dephasing times, the decay rate no longer follows a quadratic dependence, as shown in \autoref{fig:supplementary_fig_7}\,b for $T_2=1\,\text{ms}$.
Modeling the $\theta$ dependence of the decay rate for a 2D layer of nine nuclear spins that are arranged on a square grid with $2.6\,\text{\AA}$ spacing shows that $\Gamma$ is independent of $\theta$ over the investigated range. 
However, if combined with a Markovian dephasing of $T_2=1\,\text{ms}$, we once again observe an almost quadratic dependency with a flat region around $\theta/\pi=1$ as illustrated in \autoref{fig:supplementary_fig_7}\,c, which is in agreement with previous reported observations on dipolar-coupled systems~\cite{choi2017observation}.
As shown by the experimental data in \autoref{fig:supplementary_fig_7}\,c, measurement uncertainties make it difficult to distinguish genuine DTC order from Markovian dephasing.
Although the $R^2=82.81\,\%$ of the quadratic fit is relatively low, the measurement does not allow us to draw a clear conclusion on whether the system exhibits a flat region indicative of DTC order. \\
All decay rates are extracted from a fit with a stretched exponential envelope given by $a\cdot\cos(\theta\cdot N)\cdot\exp(-(\Gamma\cdot N)^n)+b$ or $a\cdot\cos(\pi\cdot N)\cdot\exp(-(\Gamma\cdot N)^n)+b$, depending on whether a beating is visible. 
We always perform both fits and chose the one with higher $R^2$.
While these fits generally yield $R^2>99\,\%$, we would like to highlight once again that this analysis/ measure is only performed for the first 40 Floquet cycles.
In the nine-spin simulation we observe revivals of DTC order at larger $N$ for short interaction times $\tau$, which initially appear as a decay for the first 40 cycles.
It remains unclear whether this behavior is intrinsic to dipolar-coupled systems or an artifact of the limited system size. 
Since our simulations reproduce most of the observed dynamics accurately and the \nuc layer also extends across the entire diamond, we expect the revivals not to result from finite-size effects.
Even if they did, their impact on our results appears negligible.
Current setup limitations prevent us from probing these dynamics at larger $N$ and make the decay rate extraction for small $\Gamma$ less accurate.
More efficient rf control, like conducting structures on the diamond surface or external coils could help to overcome these limitations in the future.
%

\subsection{Origin of Markovian Noise}

\begin{figure}
    \centering
    \includegraphics[width=0.48\textwidth]{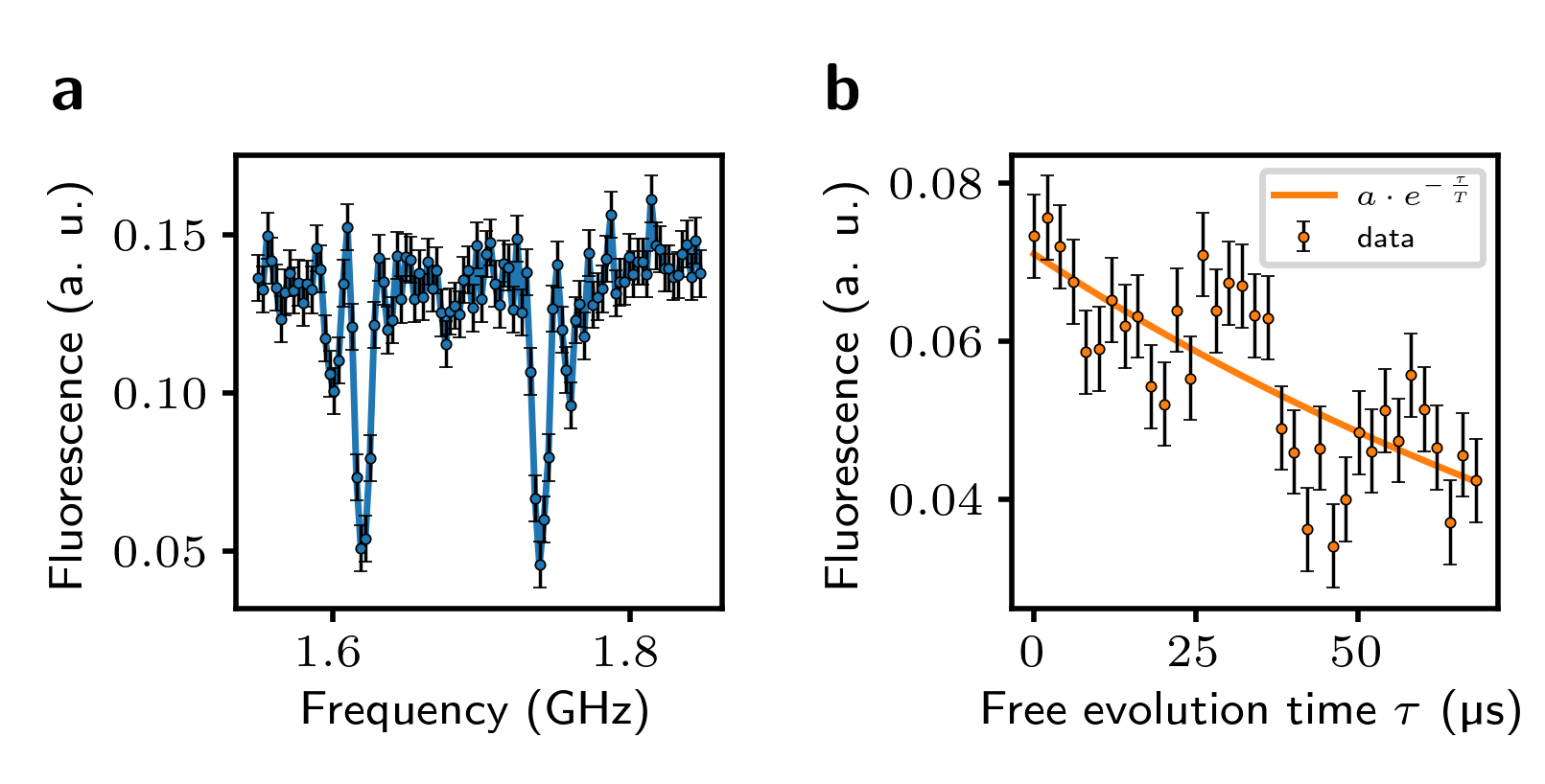}
    \caption{DEER experiments.
    \textbf{a} DEER spectrum on the \nv center, showing the distinct $\text{P}_1$ center resonances by sweeping the frequency of the $\pi$ pulse on the electron spin environment. 
    \textbf{b} Shifting the spacing between the \nv center's $\pi$ pulse and the $\text{P}_1$ center's $\pi$ pulse allows to probe the \nv center's coherence time under the associated spin noise.
    The solid line shows the exponential fit.
    }
    \label{fig:supplementary_fig_11}
\end{figure}

We can estimate the Markovian dephasing by probing the local electron spin environment via a DEER measurement on the \nv center.
The DEER sequence is equivalent to a Hahn echo sequence, with an additional $\pi$ pulse applied on the electron spin environment simultaneously with the echo's inversion $\pi$ pulse.
The surrounding $\text{P}_1$ centers, originating from implanted nitrogen atoms, are expected to be the main source of Markovian dephasing~\cite{findler2024detecting}. 
\autoref{fig:supplementary_fig_11}\,a shows the DEER spectrum obtained by sweeping the frequency of the $\pi$ pulse on the electron spin environment, revealing a clear coupling to nearby $\text{P}_1$ centers.
While the $\text{P}_1$ center is a $S=1/2$ system, strong hyperfine coupling to its intrinsic \nit nuclear spin leads to further splitting of the resonances~\cite{degen2021entanglement}. 
By sitting on a $\text{P}_1$ center resonance and varying the spacing between the \nv center’s $\pi$ pulse and the $\text{P}_1$ center’s $\pi$ pulse, allows to probe the \nv center's coherence time for the associated spin noise, as shown in \autoref{fig:supplementary_fig_11}\,b.
We observe a coherence time of $130\,\text{\textmu s}\pm20\,\text{\textmu s}$ for the \nv center, corresponding to $340\,\text{ms}\pm60\,\text{ms}$ for our \nuc nuclear spins given their $\sim 2600$ times smaller gyromagnetic ratio.
Assuming that the $\text{P}_1$ centers are the main source of Markovian dephasing for our nuclear spins, we can rule out Markovian dephasing as the origin of the observed DTC order.
%

\subsection{Markovian dephasing or DTC order?}
\label{sec:dephasing_or_order}

\begin{figure}
    \centering
    \includegraphics[width=0.48\textwidth]{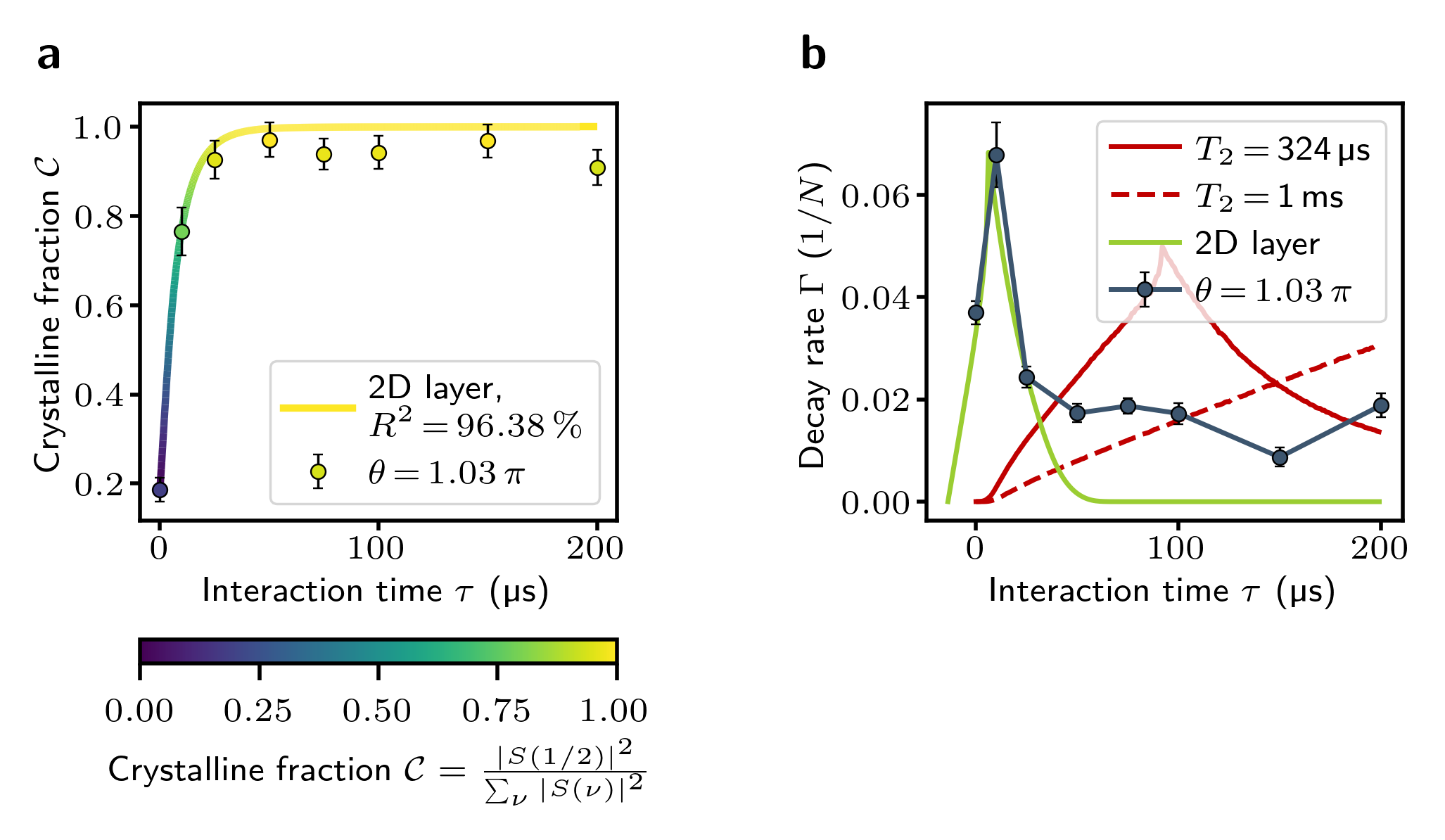}
    \caption{Crystalline fraction and decay rates.
    \textbf{a} Crystalline fraction $\mathcal{C}$ for increasing interaction times $\tau$.
    The measurement data is fit with a simulation of nine interacting nuclear spins placed on a square grid.
    \textbf{b} Decay rates as a function of the interaction time $\tau$ for $\theta=1.03\,\pi$.
    The blue points represent the measurement data, while the red lines shows a simulation of a single nuclear spin subject to Markovian dephasing.
    The green line corresponds to the simulation of a 2D layer with nine coupled nuclear spins, matching the data well at the initial measurement points.
    }
    \label{fig:supplementary_fig_14}
\end{figure}

\begin{figure}
    \centering
    \includegraphics[width=0.48\textwidth]{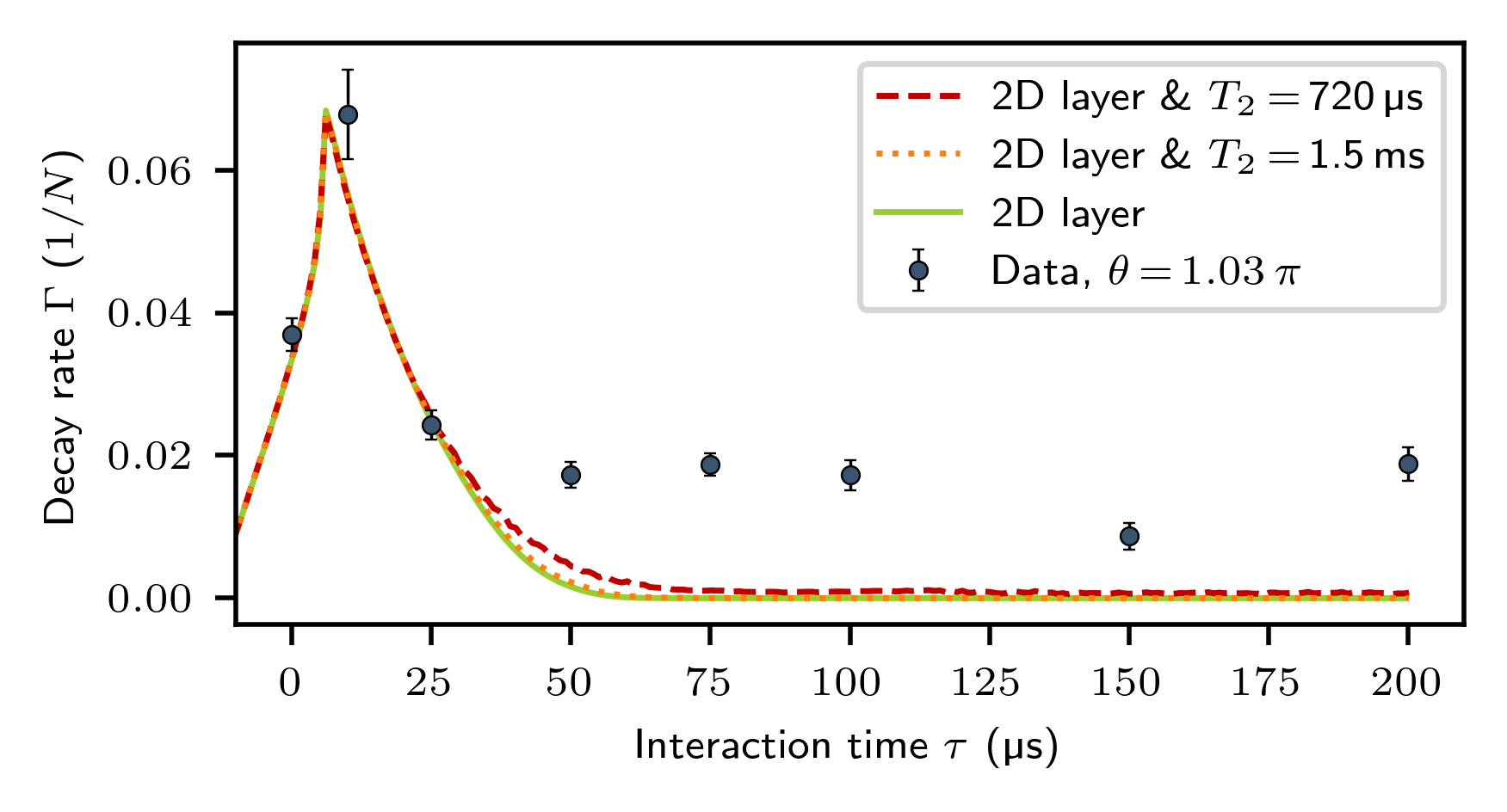}
    \caption{Decay rate simulation with Markovian dephasing.
    We simulate the DTC experiment (blue data points) with the 2D layer model of nine nuclear spins with (orange and red line) and without Markovian dephasing (green line).
    For realistic dephasing times, adding Markovian dephasing does not reproduce the observed plateau. 
    }
    \label{fig:supplementary_fig_13}
\end{figure}

As shown in \autoref{sec:influence_markovian_dephasing}, sweeping $\theta$ does not clearly reveal whether the observed stabilization of the oscillation for increasing $\tau$ originates from true DTC order or Markovian dephasing.
For a clearer distinction, we first approximate the measured crystalline fraction ($M=300$) with a simulation of nine interacting \nuc spins forming a 2D square grid.
The simulation is fitted to the measurement data by varying the distances between the spins.
To align with the diamond's growth direction, the layer is rotated by $54.7^\circ$, and the simulation data is shifted by the rotation pulse length to account for its finite duration. 
This results in a spacing of $2.6\,\text{\AA}$, corresponding to an average coupling of $(2\pi)\,172\,\text{Hz}$.
The simulation closely fits the measurement data as seen in \autoref{fig:supplementary_fig_14}\,a, yielding $R^2=96.38\,\%$. \\
Next we plot $\Gamma$ for varying $\tau$ in \autoref{fig:supplementary_fig_14}\,b.
The red line depicts the simulation of a single nuclear spin under Markovian dephasing, using the measured $T_2^*=324\,\text{\textmu s}$ as dephasing time.
Longer dephasing times flatten the lineshape and shift the peak to longer $\tau$, reducing the agreement with the measurement data.
The WAHUHA measurements of \autoref{fig:supplementary_fig_2} indicate that Markovian dephasing greatly exceeds $T_2^*$, ruling it out as the origin of the observed order. \\
The nine-spin model matches the initial data points and qualitatively reproduces the lineshape.
For $\tau>25\,\text{\textmu s}$ the measured decay rate remains approximately constant. 
Adding Markovian dephasing with reasonable dephasing times to the nine spin model does not reproduce the observed plateau of the measurement data, as shown in \autoref{fig:supplementary_fig_13}.
The Markovian dephasing appears to have little effect on the system's response. 
Instead, spin diffusion beyond the detection volume and the nuclear spins' $T_1$ process may contribute to the observation. 

\subsection{Normalization Errors}
\label{sec:normalization_errors}

\begin{figure}[h]
    \centering
    \includegraphics[width=0.49\textwidth]{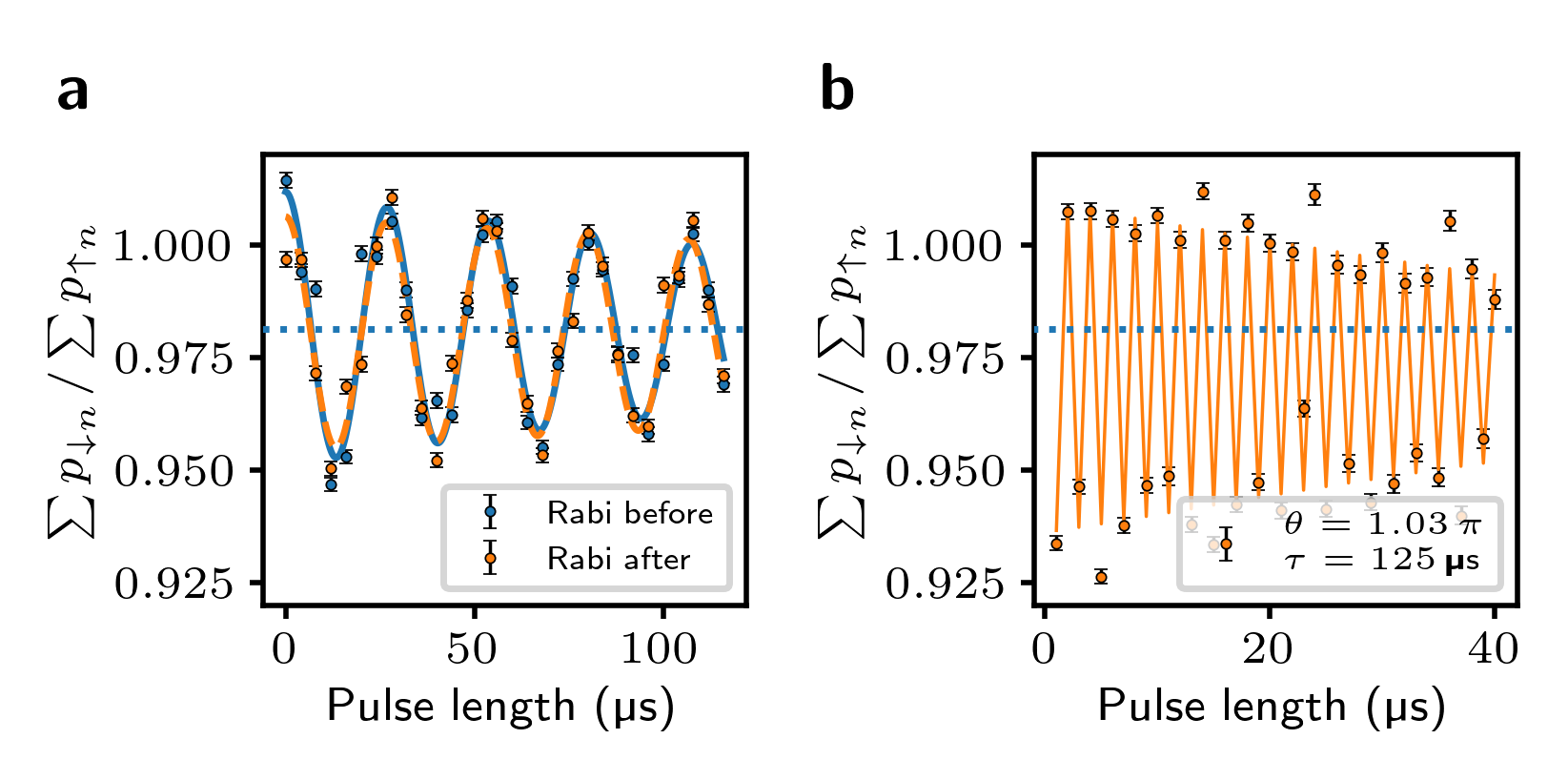}
    \caption{Normalization errors.
    \textbf{a} Nuclear Rabi measurement before (blue) and after (orange) the DTC experiments.
    The y-axis shows the ratio between the NOVEL curves according to \autoref{eq:readout}.
    The solid line represents the fits and the dotted blue line shows the corresponding baseline.
    \textbf{b} DTC experiment for a rotation angle of $\theta=1.03\,\pi$ and an interaction time $\tau=125\,\text{\textmu s}$.
    The solid orange line shows the fit result, while the dotted line indicates the Rabi fit baseline from subfigure a for comparison. 
    }
    \label{fig:supplementary_fig_10}
\end{figure}

%
\autoref{fig:supplementary_fig_10}\,a shows the nuclear Rabi measurement of the \nv center used for all presented DTC measurements.
The fit results (cosine with stretched exponential decay) are used to normalize the measurement data and infer the $\langle I_z\rangle$ expectation value according to \autoref{eq:readout_normalization}. \\
The normalization can be affected by multiple errors.
Temperature instabilities can slightly shift the magnetic field, which changes the \nv center’s transition frequency and detunes the applied pulse gates. 
This detuning modifies the effective Rabi frequency, shifting the Hartmann–Hahn resonance and altering the shape of the acquired NOVEL polarization curves.
Strong rf driving can physically displace the wire used to deliver the mw and rf fields, changing the Rabi frequency for a fixed amplitude and thereby shifting the Hartmann-Hahn resonance. \\
For our DTC experiments we observe a significant increase of the signal's amplitude, as shown in \autoref{fig:supplementary_fig_10}\,b.
This causes incorrect normalization of the measurement data and a faulty reconstruction of the $\langle I_z \rangle$ expectation value.
However, it does not affect the extraction of the decay rate nor the calculation of the crystalline fraction.
The enlarged amplitude suggests that more nuclear spins coherently contribute to the obtained signal than in the Rabi measurement.
To verify that the enlarged amplitude is not an artifact of setup instabilities, the Rabi experiment was repeated immediately after the DTC measurements (orange data points in \autoref{fig:supplementary_fig_10}\,a), yielding the same result as before.
It is unclear where this effect originates from.

\section{Experimental Setup}
\label{sec:supplementary_setup}

\begin{figure}
    \centering
    \includegraphics[width=0.48\textwidth]{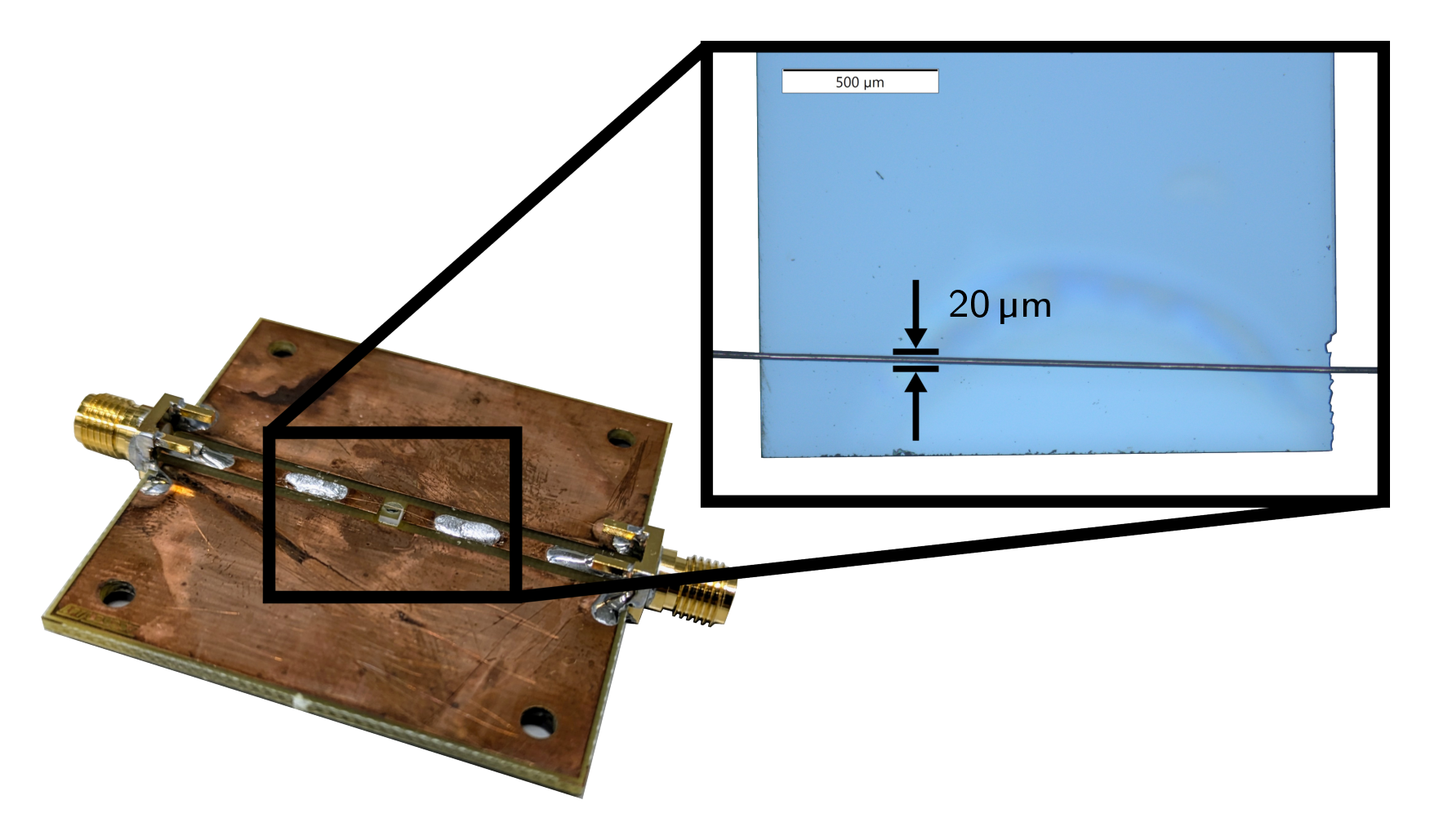}
    \caption{Diamond sample and sample holder.
    The sample holder is a copper-plated PCB with SMA connectors.
    The diamond is glued in a central gap of the signal line and a 20\,\textmu m thin copper wire is soldered across the diamond to connect the line.}
    \label{fig:supplementary_fig_8}
\end{figure}

All measurements are performed in a confocal setup at a temperature of $T=23.1\,{}^\circ\text{C} \pm 0.5\,{}^\circ\text{C}$.
The error of the temperature corresponds to the standard deviation, measured over the course of one month. \\
Due to the large amount of measurements required for our studies, we performed the experiments for sample A and B on two different, but equivalent setups inside the same laboratory. 
Measurements related to the AXY distance estimation were performed on the same setup. \\
The confocal setups feature either a fiber-coupled, pulsed 517\,nm laser (\textcolor{gray}{Toptica, iBeam smart}) or a continuous wave, 561\,nm laser (\textcolor{gray}{Coherent, Sapphire}), which is then pulsed through an AOM (\textcolor{gray}{Crystal Technology, 3200-146}) and coupled through a single-mode, polarization-maintaining fiber.
The green laser light is reflected by a beam sampler and focused through an immersion oil objective (\textcolor{gray}{Olympus, PlanApo N 6x/ 1.42\,NA}) onto the diamond.
To obtain a confocal image, we scan the objective with a nano-positioning stage (\textcolor{gray}{Physik Instrumente, P-562.3CD}).
The polarization of the incoming laser light is adjusted via a $\lambda/2$-waveplate, to match the \nv center's optical dipole axis. \\ 
The red fluorescence light is collected by the objective, passes the beam sampler and is guided through a pinhole onto an Avalanche photodiode (\textcolor{gray}{Excelitas, SPCM-AQRH-14}).
The fluorescence light is filtered by a 650\,nm longpass filter to ensure clear imaging of the \nv centers. \\  
We use a FPGA (\textcolor{gray}{OpalKelly}) to copy the photon time-trace which is then send to a multifunction I/O device (\textcolor{gray}{National Instruments, 6343}) for confocal imaging and a digitizer (\textcolor{gray}{FAST ComTec, MCS6 or home-built OpalKelly FPGA counter}) for fast, pulsed measurements. \\
Microwave and radio-frequency control fields and all trigger signals are generated by an arbitrary waveform generator (\textcolor{gray}{Keysight, M8195A or Tektronix, 70001A}).
The non-inverted analog output of the AWG is connected to a microwave amplifier (\textcolor{gray}{Amplifier Research, 30S1G3M2 or 60S1G4A}) and the inverted one to a radio-frequency amplifier (\textcolor{gray}{Amplifier Research, 25U1000 or 125A400}).
Since the microwave and radio-frequency line share the same analog output, we install a DC block and a 400\,MHz low-pass in the radio-frequency path before the amplifier to suppress unwanted noise during microwave pulse operations.
Both lines are combined by a diplexer (\textcolor{gray}{Amphenol Procom, PRO-DIPX 520/790-2700-N(f)}) and finally send to the sample holder, which is shown in \autoref{fig:supplementary_fig_8}. \\
The sample holder itself is a copper-plated PCB board equipped with SMA connectors.
The signal line features a gap, in which we glue our diamond samples.
We then solder a 20\,\textmu m thin copper wire over the diamond's surface to connect the signal line. 
The output of the sample holder is attenuated and terminated by a 50\,$\Omega$ termination. \\
Placed on a micro-positioning stage, we use permanent magnets to create the static, bias magnetic field.

\bibliography{references}